\documentclass[aps,prl,amsmath,amssymb,superscriptaddress,twocolumn,10pt,longbibliography]{revtex4-2}

\usepackage{times,txfonts}
\usepackage{nicefrac}
\usepackage{xr-hyper}
\usepackage[colorlinks=true,linkcolor=blue,urlcolor=blue,citecolor=magenta,pdfusetitle]{hyperref}
\usepackage{epsfig} 
\usepackage{subfigure}
\usepackage{physics}
\usepackage{bm}
\usepackage{xcolor}
\usepackage[normalem]{ulem}

\begin{document}

\title{Directional emission and photon bunching from a qubit pair in waveguide}

\author{Maria Maffei}

\author{Domenico Pomarico}

\author{Paolo Facchi}

\author{Giuseppe Magnifico}

\author{Saverio Pascazio}

\author{Francesco V. Pepe}

\affiliation{Dipartimento di Fisica, Universit\`{a} di Bari, I-70126 Bari, Italy}
\affiliation{INFN, Sezione di Bari, I-70125 Bari, Italy}

\date{\today}

\begin{abstract}
Waveguide quantum electrodynamics represents a powerful platform to generate entanglement and tailor photonic states. We consider a pair of identical qubits coupled to a parity invariant waveguide in the microwave domain. By working in the one- and two-excitation sectors, we provide a unified view of decay processes and we show the common origin of directional single photon emission and two photon directional bunching. Unveiling the quantum trajectories, we demonstrate that both phenomena are rooted in the selective coupling of orthogonal Bell states of the qubits with photons propagating in opposite directions. We comment on how to use this mechanism to implement optimized post-selection of Bell states, heralded by the detection of a photon on one side of the system.
\end{abstract}

\maketitle

\textit{Introduction}~-~In recent years, the field of waveguide quantum electrodynamics~\cite{onedim_review, cqed1, cqed2, onedim1, onedim2, onedim3, onedim4, onedim5, onedim6, kimble1, focused1, focused2, focused3, mirror1, mirror2, atomrefl1, yudsonPLA} has seen an endeavour towards the implementation of networks to communicate and manipulate information encoded in itinerant photons~\cite{Kimble_2008,Pichler2017,Cala-Pich2019, interferometry, multi-comm, wqed-scat}. In this context, it is crucial to achieve selective and tunable directional propagation of photons. In the optical domain, this task is easily achieved by exploiting the locking of the photon polarization with the direction of propagation in the so-called chiral waveguides~\cite{pichler2015, ramos2016, lodahl2017chiral}. In the microwave domain, where this effect cannot be exploited, destructive interference between fields emitted by a pair of identical two-level systems (qubits) has been identified as a promising strategy~\cite{refereeA1, refereeA2, waveguide_pra, baranger, baranger2013, NJP, yudson2014, laakso, Fedorov1, Fedorov2, Alexia2016, Calajo, waveguide_pra4, Guimond2020unidirectional, gasparinetti, gasparinetti2023photons, solano2023}. One-dimensional arrays of multiple emitters have been extensively investigated as well~\cite{Dinc,lalumiere2013, boundstates2017, leo4, bello, waveguide_pra3, dong, fang14, gu, guimond, paulisch, calajo15, Kockum, kimble2015, waveguide_njp, oscillators} including systems in the optical domain~\cite{ramos14, bernien}.

The most natural description of the a pair of identical emitters in a parity-invariant waveguide uses centrally symmetric and antisymmetric states of the propagating electromagnetic field~\cite{waveguide_pra, waveguide_pra2}. However, such a natural formulation does not correspond to a simple experimental detectability of the two kinds of photons, symmetric or antisymmetric, unless specific interferometric techniques are employed. On the other hand, describing the dynamics in terms of photon propagation directions gives new insights on the system physics and the possibility to implement new procedures.

\begin{figure}
    \centering
    \includegraphics[width=0.97\columnwidth]{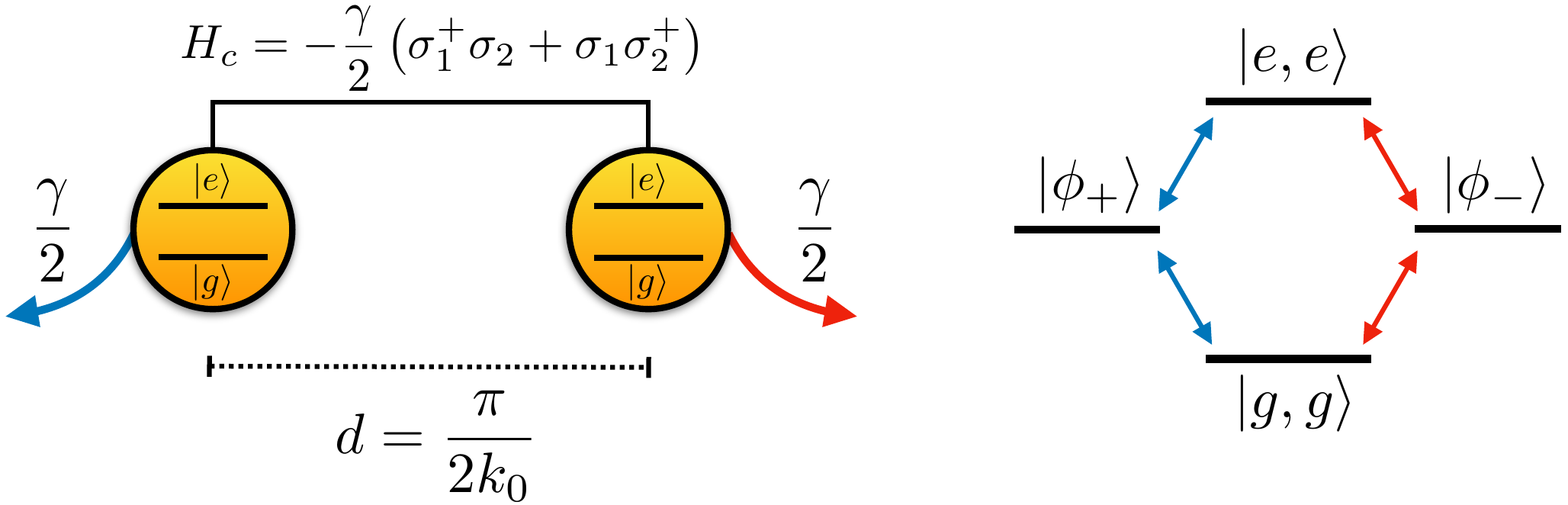}
    \caption{A pair of identical qubits of frequency $\omega_0$ are placed at a distance $d$ along a parity-invariant waveguide with linear dispersion relation and are connected among each other by a control coupling $H_c$. In the antiresonance condition, $\omega_0 \tau=k_0d=\pi/2$, with $\tau=d/v_g$ and $v_g$ being the field group velocity, the Bell state $\ket{\phi_{+}}(\ket{\phi_{-}})$ absorbs/emits only left-(right-) propagating photons. The system features an effective four-level system with optical selection rules. }
    \label{fig:panel_1}
\end{figure}

An independent emission of photons propagating to the left or to the right of the emitters can be achieved only for certain specific values of the distance between the emitters and additionally requires the implementation of a control coupling between them: two identical qubits placed a quarter wavelength apart and connected via a suitable control coupling can emit and absorb single photons directionally~\cite{Nakamura2020, kannan2023demand}. This happens as orthogonal Bell states of the qubits get coupled selectively with different photon propagation directions, see Fig.~\ref{fig:panel_1}. Remarkably the same mechanism can be used to generate two-photons N00N states~\cite{kannan2020generating,merkel2010generation,wang2011deterministic,lang2013correlations,hua2020efficient,mohseni2020deterministic,su2013fast}.

In this Letter, we provide a unified view of decay processes of a pair of qubits in the one-excitation and two-excitation sectors, showing the common origin of directional emission and bunching phenomena~\cite{yudson2008, leo3, cirac2015, paulisch2017, waveguide_pra2}. Differently from the existing theoretical literature, our results are not built on the solution of qubits master equation, but rather on that of the closed light-matter dynamics~\cite{maffei2023energy, Maffei2022Closed, Fischer}. The joint system state shows that the state of the emitted photons and their entanglement can be tuned by changing the qubits distance and the strength of the control coupling. 

The unveiling of the quantum trajectories of the joint system shows that the emission of a two-photons N00N state with directional bunching can be regarded as an avalanche process: the first photon is emitted towards left or right with equal probability hence conserving the initial parity symmetry; then, according to its direction, the qubits are projected onto a different Bell state that consequently is forced to emit the second photon in the same direction. We then show that the mechanisms underlying left/right photon emission can be used to implement optimized post-selection of Bell states, heralded by the detection of photons on one or the other side of the qubits pair.

\textit{Model and dynamics}~-~We consider a pair of identical qubits coupled to the same one-dimensional waveguide in different points, at a distance $d$ from each other. The bare Hamiltonian of qubit $j\in\{1,2\}$ is $H^{(0)}_j= \omega_0 \sigma^{\dagger}_j\sigma_j$, where $\sigma_j=|g_j\rangle\langle e_j|$, with $e$ and $g$ labelling the excited and ground state, respectively. For notation shortness, we will denote the states of the tensor product emitter basis as $\ket{ee}$, $\ket{eg}$, $\ket{ge}$, and $\ket{gg}$. The atoms are also coupled among each other directly by an energy-exchange interaction described by the Hamiltonian $H_c=  J(\sigma_1^+ \sigma_2 + \sigma_1 \sigma_2^+)$ which is called, for reasons that will be clear in the following, the \textit{cancellation coupling}. A possible strategy to implement such a term consists in buffering the emitter-waveguide coupling with interacting resonant cavities ~\cite{Nakamura2020,Guimond2020unidirectional,kannan2020generating}. The electromagnetic field propagates along the waveguide with linear dispersion relation (in the relevant bandwidth around $\omega_0$), with constant group velocity $v_g$. Hence, in the interaction picture with respect to the bare Hamiltonians of the qubits and the field, the coupling between the atoms and the waveguide photons reads, within the rotating wave approximation, 
\begin{align}\label{eq:V_continuousTime}
    V_{I}(t)=& \sqrt{\frac{\gamma}{2}}\Biggl\lbrace \sigma_{1} \left[ b^{\dagger}_{R}\left(t\right)+  b^{\dagger}_{L}\left(t\right)\right]   \\ \nonumber 
    & +\sigma_{2} \left[e^{-i \omega_0 \tau} b^{\dagger}_{R}\left(t-\tau\right)+ e^{i \omega_0 \tau} b^{\dagger}_{L}\left(t+\tau\right)\right] \Biggr\rbrace + \mathrm{H.c.}
\end{align}
Here $\tau=d/v_g$ is the time of flight between the qubits placed at $x=0$ and $x=d$, and
$b_{\ell}(t)$ with $\ell\in\lbrace R,L\rbrace$,
where $L$ and $R$ stand for left- and right-propagating photons, are the  annihilation operators (quantum noise), verifying $[b_{\ell}(t),b^{\dagger}_{m}(t')]=\delta_{\ell,m}\delta(t-t')$~\cite{gardinerzoller}. We assumed that the coupling rate $\gamma$, equal for the two propagation directions, is constant over the relevant bandwidth (first Markov approximation)~\cite{Gardiner1985Input}, with the rotating wave approximation holding true for $\gamma\ll\omega_0$ ~\cite{cohentannoudjiAPI,burgarth2023taming}.

In the following we will assume that the qubit distance $d$ has order of magnitude of the atomic wavelength. This condition implies that $\tau\sim \omega_0^{-1} \ll \gamma^{-1}$, i.e.\ the time of flight between the qubits is much smaller than the typical lifetime of their excited states. Within this regime, we can neglect the propagation delay between the qubits in the quantum noise operators replacing $b^{\dagger}_{R}(t-\tau)$ and $b^{\dagger}_{L}(t+\tau)$ with $b^{\dagger}_{R}(t)$ and $b^{\dagger}_{L}(t)$, and thus the dynamics of the two qubits can  be described by a GKLS master equation~\cite{combes_slh_2017,GKS,L}. 

Now, let us consider the quantum noise increment operators defined as 
$\mathrm{d}B_{\ell}(t)\equiv \int_{t}^{t+\mathrm{d}t} \mathrm{d}s~b_{\ell}(s)$, with $[\mathrm{d}B_{\ell}(t),\mathrm{d}B^{\dagger}_{m}(t')]=\delta_{\ell,m}\mathrm{d}t$ for $t=t'$ and $0$ otherwise,
and the associated increment of the number operator $\mathrm{d}N_{\ell}(t)\equiv \int_{t}^{t+\mathrm{d}t} \mathrm{d}s~b_{\ell}^\dag(s)b_{\ell}(s)$~\cite{gardinerzoller}. The stochastic differential equation of the unitary propagator in It\=o form reads~\cite{SM, combes_slh_2017}
\begin{align}
	\label{eq_Un}
    \mathrm{d}U(t) = \Bigg\{&\Big(-i H +\frac{1}{2}\sum_\ell \mathcal{J}_\ell^\dag \mathcal{J}_\ell  \Big)\, \mathrm{d}t      \\\nonumber
    &+\sum_\ell \left[ \mathcal{J}^{\,}_{\ell} \mathrm{d}B^{\dagger}_{\ell} - \mathcal{J}^{\dagger}_{\ell} \mathrm{d}B_{\ell} + (e^{i\omega_0\tau}-1) \,\mathrm{d}N_{\ell} \right]\Bigg\}U(t),
\end{align}
where $H=H_e+H_c$, with $H_{e}=  (\gamma/2) (\sigma_1^+ \sigma_2 + \sigma_1 \sigma_2^+)$ being an effective qubit-qubit energy-exchange interaction mediated by the electromagnetic field, and
\begin{equation} 
    \mathcal{J}_{R}  = -i \sqrt{\frac{\gamma}{2}} \left(\sigma_1 + e^{-i\omega_0\tau}\sigma_2 \right), \quad  \mathcal{J}_{L}  = -i \sqrt{\frac{\gamma}{2}} \left(\sigma_1 + e^{i\omega_0\tau} \sigma_2 \right) . \label{eq:vertexJ2}
\end{equation}
are the jump operators associated to right and left emission, respectively.

Importantly, the combination of $H_{e}$ with the cancellation coupling $H_c$ determines a new effective Hamiltonian dynamics
\begin{equation}\label{eq:Hprime}
    H = H_e + H_c =  \frac{ \gamma}{2} \left( \sin (\omega_0 \tau) - g_c \right) \left(\sigma_1^{\dagger} \sigma_2 + \sigma_1 \sigma_2^{\dagger}\right) ,
\end{equation}
with $g_c=-2J/\gamma$. Therefore, the choice $g_c=\sin(\omega_0\tau)$ cancels the exchange interaction between the emitters, $H=0$, leaving the dissipation as the only non-trivial part of the dynamics~\eqref{eq_Un} induced by the coupling with the waveguide field. 

The Schr\"odinger equation~\eqref{eq_Un} is invariant under point reflection through the center $x=d/2$, hence, in order to obtain selective directional propagation, the inherent central symmetry of the dynamics needs to be broken by preparing the system in an asymmetric initial state~\cite{Nakamura2020}. Considering the form of the light-matter coupling in~\eqref{eq_Un}, one could na\"ively expect that preparing either of the states 
\begin{equation}\label{eq:psiRL}
    \ket{\psi_L} = \frac{\ket{eg} - e^{i \omega_0 \tau} \ket{ge}}{\sqrt{2}} , \quad \ket{\psi_R} = \frac{\ket{eg} - e^{-i \omega_0 \tau} \ket{ge}}{\sqrt{2}}
\end{equation}
which are selectively annihilated by the jump operators~\eqref{eq:vertexJ2}, i.e. $\mathcal{J}_R\ket{\psi_L}=0=\mathcal{J}_L\ket{\psi_R}$, would provide pure directional emission. However, this is not the case, since the two states are generally coupled to each other by the effective Hamiltonian~\eqref{eq:Hprime}. The following analysis will show that fully directional emission occurs only in exceptional cases, characterized by specific values of emitter distance and by fine-tuned cancellation couplings. Moreover, we show that these conditions are identical to those in which two-photon emission from a doubly-excited emitter state is fully bunched in direction.

\textit{One-excitation sector}~-~Let us first consider the case where the emitters are prepared in a pure single-excitation state $\ket{\psi}=a_{eg}\ket{eg} + a_{ge}\ket{ge} $, so that the system state at a later  time $t$ reads
\begin{align}\label{eq:sp_emission}
    \ket{\Psi(t)}= & \left[ a_{eg}(t)\ket{eg} + a_{ge}(t)\ket{ge} \right] \otimes\ket{0_{R} 0_{L}} \\\nonumber 
    &+ \ket{gg} \int_{0}^{t}\mathrm{d}s \left[f_{R}(s)b_{R}^{\dagger}(s)+ f_{L}(s)b_{L}^{\dagger}(s)\right] \otimes\ket{0_{R}0_{L}},
\end{align}
where $\ket{0_R 0_L}$ is the waveguide field vacuum. The coefficients of the excited qubit states are given by the matrix elements
\begin{equation}
    a_{eg}(t)=\bra{eg} \mathcal{K}(t) \ket{\psi}, \quad a_{ge}(t)=\bra{ge} \mathcal{K}(t) \ket{\psi},
\end{equation}
of the Kraus operator $\mathcal{K}(t)\equiv \bra{0_{R} 0_{L}} U(t) \ket{0_{R} 0_{L}}$ acting on the qubits, whose analytical expression can be shown to be~\cite{SM}
\begin{equation}\label{eq_Krauss_zero}
    \mathcal{K}(t)= \left(
        \begin{matrix}
            e^{-\gamma t} & 0 & 0 & 0 \\
            0 & \frac{1}{2}\left(e^{-\frac{1}{2} \mu_{+} t} + e^{-\frac{1}{2} \mu_{-} t} \right) & \frac{1}{2}\left(e^{-\frac{1}{2} \mu_{+} t} - e^{-\frac{1}{2} \mu_{-} t} \right) & 0 \\
            0 & \frac{1}{2}\left(e^{-\frac{1}{2} \mu_{+} t} - e^{-\frac{1}{2} \mu_{-} t} \right) & \frac{1}{2}\left(e^{-\frac{1}{2} \mu_{+} t} + e^{-\frac{1}{2} \mu_{-} t} \right) & 0 \\
            0 & 0 & 0 & 1
        \end{matrix} \right) .
\end{equation}
Here, $\mu_{\pm}=\gamma_{\pm}\pm i\delta$, where
\begin{align}\label{gammadelta}
    \gamma_{\pm}=\gamma [ 1 \pm \cos(\omega_0\tau) ],~~\quad \delta = \gamma [  \sin(\omega_0\tau) - g_c ]
\end{align}
correspond respectively to the imaginary and the real part of the self-energy eigenvalues in the limit of linear dispersion relation and including the cancellation coupling, as shown in Ref.~\cite{waveguide_pra}. The single-photon amplitudes $f_{\ell}(t)$ in Eq.~\eqref{eq:sp_emission}, with $\ell\in\{R,L\}$, are given by the matrix elements
\begin{equation}\label{eq_WF1p}
    f_{\ell}(s) = \bra{gg}\mathcal{K}(t-s)\mathcal{J}_{\ell}\mathcal{K}(s)\ket{\psi},
\end{equation}
whose analytical expression is reported in the Supplemental Material~\cite{SM}.

Besides the trivial eigenvectors $\ket{ee}$ and $\ket{gg}$, the matrix $\mathcal{K}(t)$ is generally diagonalized by the real-coefficient Bell states $\ket{\psi_{\pm}} = \left(\ket{eg} \pm \ket{ge}\right)/\sqrt{2}$ with decay rates $\gamma_{\pm}$, and relative energy splitting $\delta$ determined by the Hamiltonian~\eqref{eq:Hprime}. Since they do not break central symmetry, the states $\ket{\psi_{\pm}}$ cannot give rise to any prevalence of emission in one direction. 

The most suitable candidates for directional emission would be the states~\eqref{eq:psiRL}, but, as one can observe from the form of $\mathcal{K}(t)$ in Eq.~\eqref{eq_Krauss_zero}, the dynamics generally entails transitions between them, thus hindering purely directional emission.
A remarkable exception is represented by the following cases, that we can call \textit{controlled antiresonances},
\begin{equation}\label{eq:antires_c}
    \omega_0 \tau = \left( n + \frac{1}{2} \right) \pi , \quad g_c= (-1)^n,  \quad \text{with }n\in \mathbb{N},
\end{equation}
in which the antiresonance condition on $\omega_0\tau$ makes the quantities $\gamma_{\pm}$ equal to the isolated-qubit decay rate, while the cancellation coupling is used to suppress the Hamiltonian evolution in the single-excitation sector, thus making $\mathcal{K}(t)$ diagonal. In these conditions, the right- and left-emitting states~\eqref{eq:psiRL} specialize to the \textit{orthogonal} Bell states $\ket{\phi_{\pm}} = \left(\ket{eg} \pm i \ket{ge}\right)/\sqrt{2}$: For even-$n$, resp. odd-$n$, antiresonances one finds $\ket{\psi_{R}}=\ket{\phi_+}$ and $\ket{\psi_{L}}=\ket{\phi_-}$, resp. $\ket{\psi_{R}}=\ket{\phi_-}$ and $\ket{\psi_{L}}=\ket{\phi_+}$. Due to the cancellation condition $g_c=(-1)^n$, no coherent transition between the two states $\ket{\phi_{\pm}}$ occurs, and the preparation of either of them at the initial time generates pure directional emission. In this case, the two qubits can be regarded as a 4-level system with optical selection rules~\cite{Guimond2020unidirectional}, as depicted in Fig.~\ref{fig:panel_1}(a), corresponding to an even-$n$ controlled antiresonance.

In general, the directionality of the emitted field can be quantified through the ratio $r_1(\ket{\psi}) = \mathcal{P}^{(\psi)}_{L}/\mathcal{P}^{(\psi)}_{R}$ with $\mathcal{P}^{(\psi)}_{L/R}=\int_{0}^{\infty}dt |f_{L/R}^{(\psi)}(t)|^2$ being the probability that the state $\ket{\psi}$ emits towards left/right. The states $\ket{\psi_{L/R}}$  of Eq.~\eqref{eq:psiRL} yield:
\begin{align}\label{eq:r1L}
    r_1(\ket{\psi_{L}})=\frac{1+\left(g_c -\sin{(\omega_0\tau)}\right)^2+\sin^2{(\omega_0\tau)}}{1+\left(g_c -\sin{(\omega_0\tau)}\right)^2-\sin^2{(\omega_0\tau)}},
\end{align}
and $r_1(\ket{\psi_{R}})=1/r_1(\ket{\psi_{L}})$. Hence, as expected, the emission is purely directional, i.e. $r_1(\ket{\psi_{L}})=\infty$ and $r_1(\ket{\psi_{R}})=0$, provided $\omega_0\tau$ and $g_c$ verify the controlled antiresonance condition in Eq.~\eqref{eq:antires_c}. It is interesting to compare the above result with the one obtained using initial one-excitation states that break spatial inversion symmetry but are factorized, i.e.\ $\ket{eg}$ and $\ket{ge}$. In this case, one finds   
\begin{equation}\label{eq:r1eg}
    r_1(\ket{eg})=r_1(\ket{eg})^{-1}=3-2\frac{g_c^2+\cos^2{(\omega_0\tau)}}{1+g_c\left(g_c-\sin{(\omega_0\tau)}\right)}.
\end{equation}
Therefore, despite the cancellation coupling, if the initial states are not tailored for pure directional emission, the directionality ratio can never exceed the value $r_1=3$ \cite{waveguide_pra2}. 

\begin{figure}
    \centering
    \includegraphics[width=0.75\columnwidth]{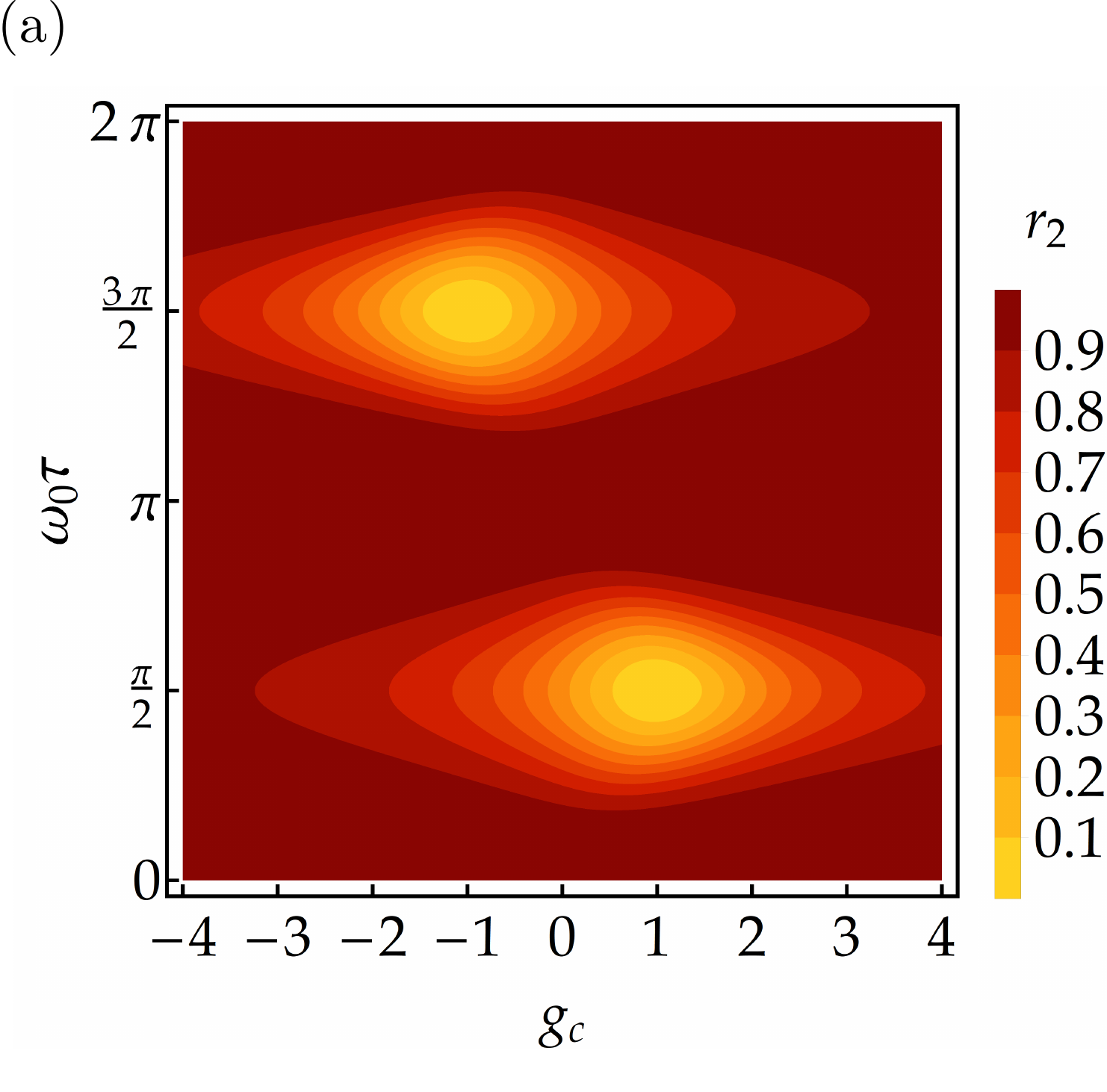}\\\includegraphics[width=0.85\columnwidth]{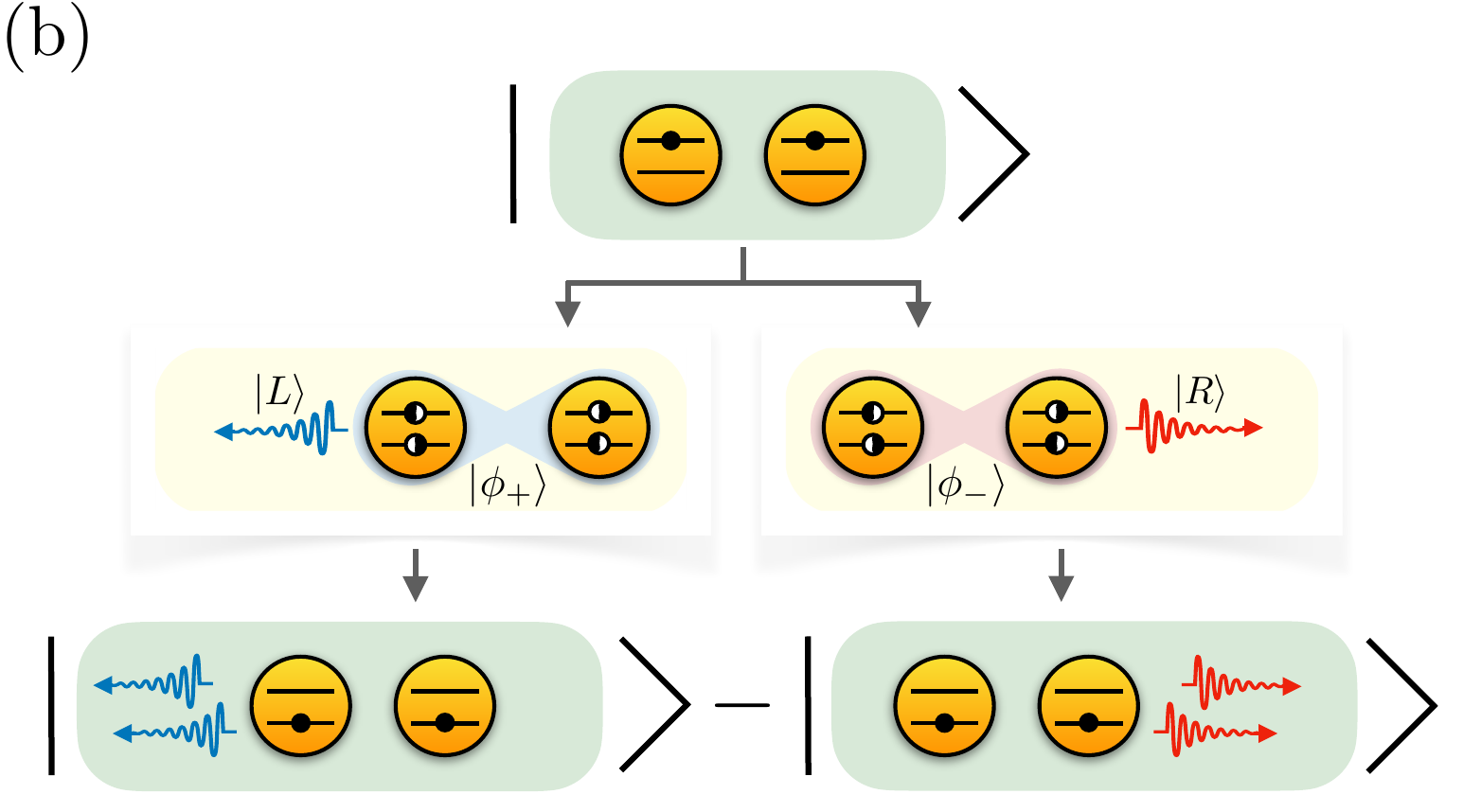}
    \caption{(a) Plot of the ratio $r_2$ in Eq.~\eqref{eq_r2}, as a function of $\omega_0\tau$ and $g_c$. As expected, $r_2=0$ (purely parallel emission) in the points verifying the even-$n$ (resp. odd-$n$) controlled antiresonce condition, $\lbrace\omega_0 \tau,g_c \rbrace =\lbrace \frac{\pi}{2},1\rbrace$ (resp. $\lbrace\omega_0 \tau,g_c \rbrace =\lbrace \frac{3\pi}{2},-1\rbrace$). (b) Pictorial representation of the quantum trajectories underlying the N00N state generation: when the controlled antiresonance condition is verified, the doubly excited state undergoes a sequence of transitions selectively coupled with the two directions of propagation, see Eq.~\eqref{eq_trajN00N}.}
    \label{fig:panel_2}
\end{figure}

\textit{Two-excitation sector: entangled photons and Bell state post-selection}~-~When the qubits are prepared in the doubly-excited state $\ket{ee}$, which is obviously centrally symmetric, the state at time $t$ comprises three different amplitudes describing: (1) both emitters remaining excited; (2) the emitters being in a single-excitation state and one photon being emitted; (3) the emitters being in the state $\ket{gg}$ and two photons being emitted. The dynamics of such an evolution is strongly influenced by what occurs in the one-excitation sector, especially concerning the alternation of resonances (i.e., the cases $\omega_0\tau=n\pi$ with $n\in\mathbb{N}$, $n\neq 0$, when one of the two Bell states $\ket{\psi_{\pm}}$ is stable) and antiresonances~\cite{waveguide_pra2}. Though the Schr\"odinger equation~\eqref{eq_Un}
enables us to determine the dynamics at any  time (see~\cite{SM}), let us focus on the asymptotic regime $t\gg \gamma^{-1}$, when only the amplitudes of type (3) survive, i.e. the emitters are found in $\ket{gg}$, and the field state reads
\begin{equation}\label{eq:sp_emission2p}
    \ket{\Xi}=\!\!\!\sum_{\ell,m\in\{R,L\}} \int_{0}^{\infty}\!\!\mathrm{d} t_2\int_{0}^{t_2}\!\!\mathrm{d} t_1 \left[ \lambda_{\ell,m}(t_1,t_2)b_{\ell}^{\dagger}(t_1)b_{m}^{\dagger}(t_2)\right]\ket{0_{R}0_{L}},
\end{equation}
with normalization achieved in the limit $t\to\infty$. The two-photon wavefunctions $\lambda_{\ell,m}(t_1,t_2)$ with $t_1\leq t_2$ are given by 
\begin{equation}\label{eq_lambda}
    \lambda_{\ell,m}(t_1,t_2)=
    \lim_{t\to\infty}\bra{gg}\mathcal{K}(t-t_2)\mathcal{J}_{m}\mathcal{K}(t_2-t_1) \mathcal{J}_{\ell}\mathcal{K}(t_1)\ket{ee}.
\end{equation}
The properties $\lambda_{R,R}(t_1,t_2)=e^{2 i \omega_0 \tau}\lambda_{L,L}(t_1,t_2)$ and $\lambda_{R,L}(t_1,t_2)=\lambda_{L,R}(t_1,t_2)$ imply that the probabilities $\mathcal{P}_{l,m}=\int_{0}^{\infty}dt_2\int_{0}^{t_2} dt_1 |\lambda_{l,m}(t_1,t_2)|^2$ are invariant under exchange of the propagation direction, as expected by the central symmetry of the initial state of the system. The ratio between the probabilities of antiparallel emission, $\mathcal{P}_{\upharpoonleft \! \downharpoonright}=\mathcal{P}_{L,R} + \mathcal{P}_{R,L}$, and that of parallel emission, $\mathcal{P}_{\parallel}=\mathcal{P}_{L,L}+\mathcal{P}_{R,R}$, is equal to:
 \begin{equation}\label{eq_r2}
    r_2 = \frac{\mathcal{P}_{\upharpoonleft \! \downharpoonright}}{\mathcal{P}_{\parallel}}= \frac{ (2 - \sin^2(\omega_0\tau)) [ 1 + (\sin(\omega_0\tau) - g_c )^2] + \sin^4(\omega_0\tau) }{ (2 - \sin^2(\omega_0\tau)) [ 1 + (\sin(\omega_0\tau) - g_c )^2] - \sin^4(\omega_0\tau) } .
\end{equation}
In the absence of cancellation coupling, $r_2$ is always comprised between the value $1/3$, reached at antiresonance, and the value $1$ reached at resonance (see the small-coupling limit in Ref.~\cite{waveguide_pra2}). Thus, in this case, antiparallel emission can never be suppressed, in accordance with the findings in the single-excitation sector. Instead, adding a cancellation coupling that verifies the controlled antiresonance condition~\eqref{eq:antires_c}, one achieves pure directional bunching of two-photon emission corresponding to a vanishing $r_2$, see Fig.~\ref{fig:panel_2}(a). In this case, $\lambda_{R,L}(t,t')=\lambda_{L,R}(t,t')=0$,
and the field asymptotically approaches the two-photon N00N state
\begin{equation}
\ket{\Xi}= \frac{\ket{2_{R} 0_{L}}-\ket{0_{R} 2_{L}}}{\sqrt{2}}.   
\end{equation}
where we have introduced the Fock state notation $\ket{n_R n_L}=(n_R!n_L!)^{-1/2} (c_R^{\dagger})^{n_R} (c_L^{\dagger})^{n_L} \ket{0_R 0_L}$ associated to the mode operators $c_{R/L}\equiv \int dt \sqrt{\gamma}e^{-\gamma t/2} b_{R/L}(t)$ \cite{Loudon1990, Combes2012PRA}, with $\ket{0_R 0_L}$ the field vacuum. Notice that, due to the phases acquired in point inversion, the above state is centrally \textit{symmetric}, as well as the initial state $\ket{ee}$.

Looking at Eq.~\eqref{eq_lambda} the entangling mechanism appears transparent. As $\mathcal{K}(t)\ket{ee}=e^{-\gamma t}\ket{ee}$ (see Eq.~\eqref{eq_Krauss_zero}), the first jump operator acts on $\ket{ee}$ and projects it onto one of the states $\ket{\phi_{\pm}}$, which is then forced to emit the second photon towards the same direction as the first one. Therefore, in the cases~\eqref{eq:antires_c}, the decay of the $\ket{ee}$ occurs with equal probability through the uncoupled channels, see Fig.~\ref{fig:panel_2}(b):
\begin{equation}\label{eq_trajN00N}
    \ket{ee} \to \left\{ \begin{array}{c}
         \ket{\psi_R}\otimes\ket{1_R0_L} \to \ket{gg}\otimes \ket{2_R 0_L} \\
         \ket{\psi_L}\otimes\ket{0_R 1_L} \to \ket{gg}\otimes \ket{0_R 2_L}
    \end{array}  \right.
\end{equation}
 with $\ket{\psi_{R}}=\ket{\phi_+}$ and $\ket{\psi_{L}}=\ket{\phi_-}$ ($\ket{\psi_{R}}=\ket{\phi_-}$ and $\ket{\psi_{L}}=\ket{\phi_+}$) for even-$n$ (odd-$n$) antiresonances.

Then if two detectors are placed on the left and on the right sides of the emitters, the observation of the first photon by the left detector, resp. the right one, occurring after an average time $(2\gamma)^{-1}$, unambiguously selects the state $\ket{\psi_L}$, resp. $\ket{\psi_R}$. The selected state does not decay until the second photon is observed after an additional average time of $\gamma^{-1}$. Therefore, the two-excitation decay, Eq.~\eqref{eq_trajN00N}, makes one of two orthogonal Bell states available, with certainty, within an average time $\gamma^{-1}$. During this time, one can think of implementing strategies to adiabatically decouple the emitters from the field hence preserving the Bell state from decaying. Triggering different operations, depending on which detector clicks, it is possible to select one of the two equiprobable system trajectories. 

Let us notice that the same post-selection scheme can give access to a Bell state also within resonance conditions. In the $n$-th resonance, only one decay channel is open: 
 \begin{equation}
    \ket{ee} \to \ket{\psi_{(-1)^n}} \otimes \ket{1_{(-1)^n}} \to \ket{gg} \otimes \ket{2},
\end{equation}
with $\ket{1_{\pm}}=(\ket{1_R,0_L}\pm\ket{0_R,1_L})/\sqrt{2}$ being one-photon states satisfying central symmetry (antisymmetry) and $\ket{2}=(\ket{2_R 0_L}+\ket{0_R 2_L}+\sqrt{2}\ket{1_R 1_L})/2$. In this case, as before, the first emitted photon is observed by one of the two detectors after an average time $(2\gamma)^{-1}$. Regardless of which detector clicks, the qubits are left in the Bell state $\ket{\psi_{+}}$, resp. $\ket{\psi_{-}}$, for an even-$n$, resp. an odd-$n$ resonance. However, such a state subsequently decays \textit{twice as faster} than $\ket{\phi_{\pm}}$ at antiresonance. This halves the time for possible operations to decouple the emitters from the field to preserve the Bell state. 

Finally, when the system is neither in a resonance nor in an antiresonance, one can still post-select one of the states $\ket{\psi_{\pm}}$, with uneven relative probabilities $\gamma_{\pm}/(2\gamma)=(1\pm\cos(\omega_0 \tau))/2$, but this would require an interferometric detection scheme to distinguish between symmetric and antisymmetric single photon states $\ket{1_{\pm}}$. Moreover, it is worth noticing that in this case the most probable state to be detected is also the one that decays faster afterwards.

\textit{Outlook}~-~We presented an analytic description of the dynamics underlying the directional emission of single-photons and the generation of two-photons N00N states from a pair of qubits in waveguide. We displayed the common root of these phenomena emerging in the same antiresonance conditions, hence highlighting the primary role played by central symmetry. The proposed approach emerges as the ideal candidate to achieve exact modeling and characterization of arrays of multiple qubits and multi-level systems (qudits) whose dynamics and collective properties are determined by the symmetries, in particular in the microwave domain~\cite{interferometry}. Furthermore, by describing the closed-system dynamics through a collision model, our analysis can be extended in order to include a proper description of time delays, feedback~\cite{ZollerCM, guimond} and scattering phenomena~\cite{Calajo_scattering, guimond2017scattering}.

\textit{Acknowledgement}~-~We thank G. Calaj\`o for inspiring discussions. We acknowledge the support by Regione Puglia and QuantERA ERA-NET Cofund in Quantum Technologies (GA No.\ 731473), project PACE-IN, and by INFN through the project ``QUANTUM''. 
PF acknowledges the support by the Italian National Group of Mathematical Physics (GNFM-INdAM). PF and GM acknowledge the support by PNRR MUR project CN00000013-``Italian National Centre on HPC, Big Data and Quantum Computing''. MM, SP and FVP acknowledge the support by PNRR MUR project PE0000023-NQSTI.


\begin{thebibliography}{91}%
\makeatletter
\providecommand \@ifxundefined [1]{%
 \@ifx{#1\undefined}
}%
\providecommand \@ifnum [1]{%
 \ifnum #1\expandafter \@firstoftwo
 \else \expandafter \@secondoftwo
 \fi
}%
\providecommand \@ifx [1]{%
 \ifx #1\expandafter \@firstoftwo
 \else \expandafter \@secondoftwo
 \fi
}%
\providecommand \natexlab [1]{#1}%
\providecommand \enquote  [1]{``#1''}%
\providecommand \bibnamefont  [1]{#1}%
\providecommand \bibfnamefont [1]{#1}%
\providecommand \citenamefont [1]{#1}%
\providecommand \href@noop [0]{\@secondoftwo}%
\providecommand \href [0]{\begingroup \@sanitize@url \@href}%
\providecommand \@href[1]{\@@startlink{#1}\@@href}%
\providecommand \@@href[1]{\endgroup#1\@@endlink}%
\providecommand \@sanitize@url [0]{\catcode `\\12\catcode `\$12\catcode `\&12\catcode `\#12\catcode `\^12\catcode `\_12\catcode `\%12\relax}%
\providecommand \@@startlink[1]{}%
\providecommand \@@endlink[0]{}%
\providecommand \url  [0]{\begingroup\@sanitize@url \@url }%
\providecommand \@url [1]{\endgroup\@href {#1}{\urlprefix }}%
\providecommand \urlprefix  [0]{URL }%
\providecommand \Eprint [0]{\href }%
\providecommand \doibase [0]{https://doi.org/}%
\providecommand \selectlanguage [0]{\@gobble}%
\providecommand \bibinfo  [0]{\@secondoftwo}%
\providecommand \bibfield  [0]{\@secondoftwo}%
\providecommand \translation [1]{[#1]}%
\providecommand \BibitemOpen [0]{}%
\providecommand \bibitemStop [0]{}%
\providecommand \bibitemNoStop [0]{.\EOS\space}%
\providecommand \EOS [0]{\spacefactor3000\relax}%
\providecommand \BibitemShut  [1]{\csname bibitem#1\endcsname}%
\let\auto@bib@innerbib\@empty
\bibitem [{\citenamefont {Roy}\ and\ \citenamefont {O.}(2017)}]{onedim_review}%
  \BibitemOpen
  \bibfield  {author} {\bibinfo {author} {\bibfnamefont {C.~M.}\ \bibnamefont {Roy}, \bibfnamefont {D.~Wilson}}\ and\ \bibinfo {author} {\bibfnamefont {F.}~\bibnamefont {O.}},\ }\bibfield  {title} {\bibinfo {title} {{Colloquium: Strongly interacting photons in one-dimensional continuum}},\ }\href@noop {} {\bibfield  {journal} {\bibinfo  {journal} {Rev. Mod. Phys.}\ }\textbf {\bibinfo {volume} {89}},\ \bibinfo {pages} {021001} (\bibinfo {year} {2017})}\BibitemShut {NoStop}%
\bibitem [{\citenamefont {Blais}\ \emph {et~al.}(2004)\citenamefont {Blais}, \citenamefont {Huang}, \citenamefont {Girvin},\ and\ \citenamefont {Schoelkopf}}]{cqed1}%
  \BibitemOpen
  \bibfield  {author} {\bibinfo {author} {\bibfnamefont {A.}~\bibnamefont {Blais}}, \bibinfo {author} {\bibfnamefont {A.}~\bibnamefont {Huang}, \bibfnamefont {R.~S.and~Wallraff}}, \bibinfo {author} {\bibfnamefont {S.~M.}\ \bibnamefont {Girvin}},\ and\ \bibinfo {author} {\bibfnamefont {R.~J.}\ \bibnamefont {Schoelkopf}},\ }\bibfield  {title} {\bibinfo {title} {{C}avity quantum electrodynamics for superconducting electrical circuits: An architecture for quantum computation},\ }\href@noop {} {\bibfield  {journal} {\bibinfo  {journal} {Phys. Rev. A}\ }\textbf {\bibinfo {volume} {69}},\ \bibinfo {pages} {062320} (\bibinfo {year} {2004})}\BibitemShut {NoStop}%
\bibitem [{\citenamefont {You}\ and\ \citenamefont {Nori}(2011)}]{cqed2}%
  \BibitemOpen
  \bibfield  {author} {\bibinfo {author} {\bibfnamefont {J.~Q.}\ \bibnamefont {You}}\ and\ \bibinfo {author} {\bibfnamefont {F.}~\bibnamefont {Nori}},\ }\bibfield  {title} {\bibinfo {title} {{A}tomic physics and quantum optics using superconducting circuits},\ }\href@noop {} {\bibfield  {journal} {\bibinfo  {journal} {Nature}\ }\textbf {\bibinfo {volume} {474}},\ \bibinfo {pages} {589} (\bibinfo {year} {2011})}\BibitemShut {NoStop}%
\bibitem [{\citenamefont {Faraon}\ \emph {et~al.}(2007)\citenamefont {Faraon}, \citenamefont {Waks}, \citenamefont {Englund}, \citenamefont {Fushman},\ and\ \citenamefont {Vu\v{c}kovi\'c}}]{onedim1}%
  \BibitemOpen
  \bibfield  {author} {\bibinfo {author} {\bibfnamefont {A.}~\bibnamefont {Faraon}}, \bibinfo {author} {\bibfnamefont {E.}~\bibnamefont {Waks}}, \bibinfo {author} {\bibfnamefont {D.}~\bibnamefont {Englund}}, \bibinfo {author} {\bibfnamefont {I.}~\bibnamefont {Fushman}},\ and\ \bibinfo {author} {\bibfnamefont {J.}~\bibnamefont {Vu\v{c}kovi\'c}},\ }\bibfield  {title} {\bibinfo {title} {{E}fficient photonic crystal cavity-waveguide couplers},\ }\href@noop {} {\bibfield  {journal} {\bibinfo  {journal} {Appl. Phys. Lett.}\ }\textbf {\bibinfo {volume} {90}},\ \bibinfo {pages} {073102} (\bibinfo {year} {2007})}\BibitemShut {NoStop}%
\bibitem [{\citenamefont {Dayan}\ \emph {et~al.}(2008)\citenamefont {Dayan}, \citenamefont {Parkins}, \citenamefont {Aoki}, \citenamefont {Ostby}, \citenamefont {Vahala},\ and\ \citenamefont {Kimble}}]{onedim2}%
  \BibitemOpen
  \bibfield  {author} {\bibinfo {author} {\bibfnamefont {B.}~\bibnamefont {Dayan}}, \bibinfo {author} {\bibfnamefont {A.~S.}\ \bibnamefont {Parkins}}, \bibinfo {author} {\bibfnamefont {T.}~\bibnamefont {Aoki}}, \bibinfo {author} {\bibfnamefont {E.~P.}\ \bibnamefont {Ostby}}, \bibinfo {author} {\bibfnamefont {K.~J.}\ \bibnamefont {Vahala}},\ and\ \bibinfo {author} {\bibfnamefont {H.~J.}\ \bibnamefont {Kimble}},\ }\bibfield  {title} {\bibinfo {title} {{A} photon turnstile dynamically regulated by one atom},\ }\href@noop {} {\bibfield  {journal} {\bibinfo  {journal} {Science}\ }\textbf {\bibinfo {volume} {319}},\ \bibinfo {pages} {1062} (\bibinfo {year} {2008})}\BibitemShut {NoStop}%
\bibitem [{\citenamefont {Vetsch}\ \emph {et~al.}(2010)\citenamefont {Vetsch}, \citenamefont {Reitz}, \citenamefont {Sague}, \citenamefont {Schmidt}, \citenamefont {Dawkins},\ and\ \citenamefont {Rauschenbeutel}}]{onedim3}%
  \BibitemOpen
  \bibfield  {author} {\bibinfo {author} {\bibfnamefont {E.}~\bibnamefont {Vetsch}}, \bibinfo {author} {\bibfnamefont {D.}~\bibnamefont {Reitz}}, \bibinfo {author} {\bibfnamefont {G.}~\bibnamefont {Sague}}, \bibinfo {author} {\bibfnamefont {R.}~\bibnamefont {Schmidt}}, \bibinfo {author} {\bibfnamefont {S.~T.}\ \bibnamefont {Dawkins}},\ and\ \bibinfo {author} {\bibfnamefont {A.}~\bibnamefont {Rauschenbeutel}},\ }\bibfield  {title} {\bibinfo {title} {{Optical Interface Created by Laser-Cooled Atoms Trapped in the Evanescent Field Surrounding an Optical Nanofiber}},\ }\href@noop {} {\bibfield  {journal} {\bibinfo  {journal} {Phys. Rev. Lett.}\ }\textbf {\bibinfo {volume} {104}},\ \bibinfo {pages} {203603} (\bibinfo {year} {2010})}\BibitemShut {NoStop}%
\bibitem [{\citenamefont {Bajcsy}\ \emph {et~al.}(2009)\citenamefont {Bajcsy}, \citenamefont {Hofferberth}, \citenamefont {Balic}, \citenamefont {Peyronel}, \citenamefont {Hafezi}, \citenamefont {Zibrov}, \citenamefont {Vuletic},\ and\ \citenamefont {Lukin}}]{onedim4}%
  \BibitemOpen
  \bibfield  {author} {\bibinfo {author} {\bibfnamefont {M.}~\bibnamefont {Bajcsy}}, \bibinfo {author} {\bibfnamefont {S.}~\bibnamefont {Hofferberth}}, \bibinfo {author} {\bibfnamefont {V.}~\bibnamefont {Balic}}, \bibinfo {author} {\bibfnamefont {T.}~\bibnamefont {Peyronel}}, \bibinfo {author} {\bibfnamefont {M.}~\bibnamefont {Hafezi}}, \bibinfo {author} {\bibfnamefont {A.~S.}\ \bibnamefont {Zibrov}}, \bibinfo {author} {\bibfnamefont {V.}~\bibnamefont {Vuletic}},\ and\ \bibinfo {author} {\bibfnamefont {M.~D.}\ \bibnamefont {Lukin}},\ }\bibfield  {title} {\bibinfo {title} {{Efficient All-Optical Switching Using Slow Light within a Hollow Fiber}},\ }\href@noop {} {\bibfield  {journal} {\bibinfo  {journal} {Phys. Rev. Lett.}\ }\textbf {\bibinfo {volume} {102}},\ \bibinfo {pages} {203902} (\bibinfo {year} {2009})}\BibitemShut {NoStop}%
\bibitem [{\citenamefont {Wallraff}\ \emph {et~al.}(2004)\citenamefont {Wallraff}, \citenamefont {Schuster}, \citenamefont {Blais}, \citenamefont {Frunzio}, \citenamefont {Huang}, \citenamefont {Majer}, \citenamefont {Kumar}, \citenamefont {Girvin},\ and\ \citenamefont {Schoelkopf}}]{onedim5}%
  \BibitemOpen
  \bibfield  {author} {\bibinfo {author} {\bibfnamefont {A.}~\bibnamefont {Wallraff}}, \bibinfo {author} {\bibfnamefont {D.~I.}\ \bibnamefont {Schuster}}, \bibinfo {author} {\bibfnamefont {A.}~\bibnamefont {Blais}}, \bibinfo {author} {\bibfnamefont {L.}~\bibnamefont {Frunzio}}, \bibinfo {author} {\bibfnamefont {R.~S.}\ \bibnamefont {Huang}}, \bibinfo {author} {\bibfnamefont {J.}~\bibnamefont {Majer}}, \bibinfo {author} {\bibfnamefont {S.}~\bibnamefont {Kumar}}, \bibinfo {author} {\bibfnamefont {S.~M.}\ \bibnamefont {Girvin}},\ and\ \bibinfo {author} {\bibfnamefont {R.~J.}\ \bibnamefont {Schoelkopf}},\ }\bibfield  {title} {\bibinfo {title} {Strong coupling of a single photon to a superconducting qubit using circuit quantum electrodynamics},\ }\href@noop {} {\bibfield  {journal} {\bibinfo  {journal} {Nature}\ }\textbf {\bibinfo {volume} {431}},\ \bibinfo {pages} {162} (\bibinfo {year} {2004})}\BibitemShut {NoStop}%
\bibitem [{\citenamefont {Astafiev}\ \emph {et~al.}(2010)\citenamefont {Astafiev}, \citenamefont {Zagoskin}, \citenamefont {Abdumalikov}, \citenamefont {Pashkin}, \citenamefont {Yamamoto}, \citenamefont {Inomata}, \citenamefont {Nakamura},\ and\ \citenamefont {Tsai}}]{onedim6}%
  \BibitemOpen
  \bibfield  {author} {\bibinfo {author} {\bibfnamefont {O.}~\bibnamefont {Astafiev}}, \bibinfo {author} {\bibfnamefont {A.~M.}\ \bibnamefont {Zagoskin}}, \bibinfo {author} {\bibfnamefont {A.~A.}\ \bibnamefont {Abdumalikov}}, \bibinfo {author} {\bibfnamefont {Y.~A.}\ \bibnamefont {Pashkin}}, \bibinfo {author} {\bibfnamefont {T.}~\bibnamefont {Yamamoto}}, \bibinfo {author} {\bibfnamefont {K.}~\bibnamefont {Inomata}}, \bibinfo {author} {\bibfnamefont {Y.}~\bibnamefont {Nakamura}},\ and\ \bibinfo {author} {\bibfnamefont {J.~S.}\ \bibnamefont {Tsai}},\ }\bibfield  {title} {\bibinfo {title} {{Resonance Fluorescence of a Single Artificial Atom}},\ }\href@noop {} {\bibfield  {journal} {\bibinfo  {journal} {Science}\ }\textbf {\bibinfo {volume} {327}},\ \bibinfo {pages} {840} (\bibinfo {year} {2010})}\BibitemShut {NoStop}%
\bibitem [{\citenamefont {Douglas}\ \emph {et~al.}(2015)\citenamefont {Douglas}, \citenamefont {Habibian}, \citenamefont {Hung}, \citenamefont {Gorshkov}, \citenamefont {Kimble},\ and\ \citenamefont {Chang}}]{kimble1}%
  \BibitemOpen
  \bibfield  {author} {\bibinfo {author} {\bibfnamefont {J.~S.}\ \bibnamefont {Douglas}}, \bibinfo {author} {\bibfnamefont {H.}~\bibnamefont {Habibian}}, \bibinfo {author} {\bibfnamefont {C.~L.}\ \bibnamefont {Hung}}, \bibinfo {author} {\bibfnamefont {A.~V.}\ \bibnamefont {Gorshkov}}, \bibinfo {author} {\bibfnamefont {H.~J.}\ \bibnamefont {Kimble}},\ and\ \bibinfo {author} {\bibfnamefont {D.~E.}\ \bibnamefont {Chang}},\ }\bibfield  {title} {\bibinfo {title} {{Q}uantum many-body models with cold atoms coupled to photonic crystals},\ }\href@noop {} {\bibfield  {journal} {\bibinfo  {journal} {Nat. Photonics}\ }\textbf {\bibinfo {volume} {9}},\ \bibinfo {pages} {326} (\bibinfo {year} {2015})}\BibitemShut {NoStop}%
\bibitem [{\citenamefont {Dorner}\ and\ \citenamefont {Zoller}(2002)}]{focused1}%
  \BibitemOpen
  \bibfield  {author} {\bibinfo {author} {\bibfnamefont {U.}~\bibnamefont {Dorner}}\ and\ \bibinfo {author} {\bibfnamefont {P.}~\bibnamefont {Zoller}},\ }\bibfield  {title} {\bibinfo {title} {Laser-driven atoms in half-cavities},\ }\href@noop {} {\bibfield  {journal} {\bibinfo  {journal} {Phys. Rev. A}\ }\textbf {\bibinfo {volume} {66}},\ \bibinfo {pages} {023816} (\bibinfo {year} {2002})}\BibitemShut {NoStop}%
\bibitem [{\citenamefont {Zumofen}\ \emph {et~al.}(2008)\citenamefont {Zumofen}, \citenamefont {Mojarad}, \citenamefont {Sandoghdar},\ and\ \citenamefont {Agio}}]{focused2}%
  \BibitemOpen
  \bibfield  {author} {\bibinfo {author} {\bibfnamefont {G.}~\bibnamefont {Zumofen}}, \bibinfo {author} {\bibfnamefont {N.~M.}\ \bibnamefont {Mojarad}}, \bibinfo {author} {\bibfnamefont {V.}~\bibnamefont {Sandoghdar}},\ and\ \bibinfo {author} {\bibfnamefont {M.}~\bibnamefont {Agio}},\ }\bibfield  {title} {\bibinfo {title} {{Perfect Reflection of Light by an Oscillating Dipole}},\ }\href@noop {} {\bibfield  {journal} {\bibinfo  {journal} {Phys. Rev. Lett.}\ }\textbf {\bibinfo {volume} {101}},\ \bibinfo {pages} {180404} (\bibinfo {year} {2008})}\BibitemShut {NoStop}%
\bibitem [{\citenamefont {Lindlein}\ \emph {et~al.}(2007)\citenamefont {Lindlein}, \citenamefont {Maiwald}, \citenamefont {Konermann}, \citenamefont {Sondermann}, \citenamefont {Peschel},\ and\ \citenamefont {Leuchs}}]{focused3}%
  \BibitemOpen
  \bibfield  {author} {\bibinfo {author} {\bibfnamefont {N.}~\bibnamefont {Lindlein}}, \bibinfo {author} {\bibfnamefont {R.}~\bibnamefont {Maiwald}}, \bibinfo {author} {\bibfnamefont {H.}~\bibnamefont {Konermann}}, \bibinfo {author} {\bibfnamefont {M.}~\bibnamefont {Sondermann}}, \bibinfo {author} {\bibfnamefont {U.}~\bibnamefont {Peschel}},\ and\ \bibinfo {author} {\bibfnamefont {G.}~\bibnamefont {Leuchs}},\ }\bibfield  {title} {\bibinfo {title} {{A} new $4\pi$ geometry optimized for focusing on an atom with a dipole-like radiation pattern},\ }\href@noop {} {\bibfield  {journal} {\bibinfo  {journal} {Laser Phys.}\ }\textbf {\bibinfo {volume} {17}},\ \bibinfo {pages} {927} (\bibinfo {year} {2007})}\BibitemShut {NoStop}%
\bibitem [{\citenamefont {Dong}\ \emph {et~al.}(2009)\citenamefont {Dong}, \citenamefont {Gong}, \citenamefont {Ian}, \citenamefont {Zhou},\ and\ \citenamefont {Sun}}]{mirror1}%
  \BibitemOpen
  \bibfield  {author} {\bibinfo {author} {\bibfnamefont {H.}~\bibnamefont {Dong}}, \bibinfo {author} {\bibfnamefont {Z.~R.}\ \bibnamefont {Gong}}, \bibinfo {author} {\bibfnamefont {H.}~\bibnamefont {Ian}}, \bibinfo {author} {\bibfnamefont {L.}~\bibnamefont {Zhou}},\ and\ \bibinfo {author} {\bibfnamefont {C.~P.}\ \bibnamefont {Sun}},\ }\bibfield  {title} {\bibinfo {title} {{I}ntrinsic cavity {QED} and emergent quasinormal modes for a single photon},\ }\href@noop {} {\bibfield  {journal} {\bibinfo  {journal} {Phys. Rev. A}\ }\textbf {\bibinfo {volume} {79}},\ \bibinfo {pages} {063847} (\bibinfo {year} {2009})}\BibitemShut {NoStop}%
\bibitem [{\citenamefont {Tufarelli}\ \emph {et~al.}(2013)\citenamefont {Tufarelli}, \citenamefont {Ciccarello},\ and\ \citenamefont {Kim}}]{mirror2}%
  \BibitemOpen
  \bibfield  {author} {\bibinfo {author} {\bibfnamefont {T.}~\bibnamefont {Tufarelli}}, \bibinfo {author} {\bibfnamefont {F.}~\bibnamefont {Ciccarello}},\ and\ \bibinfo {author} {\bibfnamefont {M.~S.}\ \bibnamefont {Kim}},\ }\bibfield  {title} {\bibinfo {title} {Dynamics of spontaneous emission in a single-end photonic waveguide},\ }\href@noop {} {\bibfield  {journal} {\bibinfo  {journal} {Phys. Rev. A}\ }\textbf {\bibinfo {volume} {87}},\ \bibinfo {pages} {013820} (\bibinfo {year} {2013})}\BibitemShut {NoStop}%
\bibitem [{\citenamefont {Shen}\ and\ \citenamefont {Fan}(2005)}]{atomrefl1}%
  \BibitemOpen
  \bibfield  {author} {\bibinfo {author} {\bibfnamefont {J.~T.}\ \bibnamefont {Shen}}\ and\ \bibinfo {author} {\bibfnamefont {S.}~\bibnamefont {Fan}},\ }\bibfield  {title} {\bibinfo {title} {{Coherent Single Photon Transport in a One-Dimensional Waveguide Coupled with Superconducting Quantum Bits}},\ }\href@noop {} {\bibfield  {journal} {\bibinfo  {journal} {Phys. Rev. Lett.}\ }\textbf {\bibinfo {volume} {95}},\ \bibinfo {pages} {213001} (\bibinfo {year} {2005})}\BibitemShut {NoStop}%
\bibitem [{\citenamefont {Yudson}(1988)}]{yudsonPLA}%
  \BibitemOpen
  \bibfield  {author} {\bibinfo {author} {\bibfnamefont {V.~I.}\ \bibnamefont {Yudson}},\ }\bibfield  {title} {\bibinfo {title} {Dynamics of the integrable one-dimensional system ``photons + two-level atoms''},\ }\href@noop {} {\bibfield  {journal} {\bibinfo  {journal} {Phys. Lett. A}\ }\textbf {\bibinfo {volume} {129}},\ \bibinfo {pages} {17} (\bibinfo {year} {1988})}\BibitemShut {NoStop}%
\bibitem [{\citenamefont {Kimble}(2008)}]{Kimble_2008}%
  \BibitemOpen
  \bibfield  {author} {\bibinfo {author} {\bibfnamefont {H.~J.}\ \bibnamefont {Kimble}},\ }\bibfield  {title} {\bibinfo {title} {The quantum internet},\ }\href@noop {} {\bibfield  {journal} {\bibinfo  {journal} {Nature}\ }\textbf {\bibinfo {volume} {453}},\ \bibinfo {pages} {1023–1030} (\bibinfo {year} {2008})}\BibitemShut {NoStop}%
\bibitem [{\citenamefont {Pichler}\ \emph {et~al.}(2017)\citenamefont {Pichler}, \citenamefont {Choi}, \citenamefont {Zoller},\ and\ \citenamefont {Lukin}}]{Pichler2017}%
  \BibitemOpen
  \bibfield  {author} {\bibinfo {author} {\bibfnamefont {H.}~\bibnamefont {Pichler}}, \bibinfo {author} {\bibfnamefont {S.}~\bibnamefont {Choi}}, \bibinfo {author} {\bibfnamefont {P.}~\bibnamefont {Zoller}},\ and\ \bibinfo {author} {\bibfnamefont {M.~D.}\ \bibnamefont {Lukin}},\ }\bibfield  {title} {\bibinfo {title} {Universal photonic quantum computation via time-delayed feedback},\ }\href@noop {} {\bibfield  {journal} {\bibinfo  {journal} {Proceedings of the National Academy of Sciences}\ }\textbf {\bibinfo {volume} {114}},\ \bibinfo {pages} {11362} (\bibinfo {year} {2017})}\BibitemShut {NoStop}%
\bibitem [{\citenamefont {Calajò}\ \emph {et~al.}(2019)\citenamefont {Calajò}, \citenamefont {Schuetz}, \citenamefont {Pichler}, \citenamefont {Lukin}, \citenamefont {Schneeweiss}, \citenamefont {Volz},\ and\ \citenamefont {Rabl}}]{Cala-Pich2019}%
  \BibitemOpen
  \bibfield  {author} {\bibinfo {author} {\bibfnamefont {G.}~\bibnamefont {Calajò}}, \bibinfo {author} {\bibfnamefont {M.~J.~A.}\ \bibnamefont {Schuetz}}, \bibinfo {author} {\bibfnamefont {H.}~\bibnamefont {Pichler}}, \bibinfo {author} {\bibfnamefont {M.~D.}\ \bibnamefont {Lukin}}, \bibinfo {author} {\bibfnamefont {P.}~\bibnamefont {Schneeweiss}}, \bibinfo {author} {\bibfnamefont {J.}~\bibnamefont {Volz}},\ and\ \bibinfo {author} {\bibfnamefont {P.}~\bibnamefont {Rabl}},\ }\bibfield  {title} {\bibinfo {title} {{Q}uantum acousto-optic control of light-matter interactions in nanophotonic networks},\ }\href@noop {} {\bibfield  {journal} {\bibinfo  {journal} {Phys. Rev. A}\ }\textbf {\bibinfo {volume} {99}},\ \bibinfo {pages} {053852} (\bibinfo {year} {2019})}\BibitemShut {NoStop}%
\bibitem [{\citenamefont {Narla}\ \emph {et~al.}(2016)\citenamefont {Narla}, \citenamefont {Shankar}, \citenamefont {Hatridge}, \citenamefont {Leghtas}, \citenamefont {Sliwa}, \citenamefont {Zalys-Geller}, \citenamefont {Mundhada}, \citenamefont {Pfaff}, \citenamefont {Frunzio}, \citenamefont {Schoelkopf},\ and\ \citenamefont {Devoret}}]{interferometry}%
  \BibitemOpen
  \bibfield  {author} {\bibinfo {author} {\bibfnamefont {A.}~\bibnamefont {Narla}}, \bibinfo {author} {\bibfnamefont {S.}~\bibnamefont {Shankar}}, \bibinfo {author} {\bibfnamefont {M.}~\bibnamefont {Hatridge}}, \bibinfo {author} {\bibfnamefont {Z.}~\bibnamefont {Leghtas}}, \bibinfo {author} {\bibfnamefont {K.~M.}\ \bibnamefont {Sliwa}}, \bibinfo {author} {\bibfnamefont {E.}~\bibnamefont {Zalys-Geller}}, \bibinfo {author} {\bibfnamefont {S.~O.}\ \bibnamefont {Mundhada}}, \bibinfo {author} {\bibfnamefont {W.}~\bibnamefont {Pfaff}}, \bibinfo {author} {\bibfnamefont {L.}~\bibnamefont {Frunzio}}, \bibinfo {author} {\bibfnamefont {R.~J.}\ \bibnamefont {Schoelkopf}},\ and\ \bibinfo {author} {\bibfnamefont {M.~H.}\ \bibnamefont {Devoret}},\ }\bibfield  {title} {\bibinfo {title} {{Robust Concurrent Remote Entanglement Between Two Superconducting Qubits}},\ }\href@noop {} {\bibfield  {journal} {\bibinfo  {journal} {Phys. Rev. X}\ }\textbf {\bibinfo {volume} {6}},\ \bibinfo {pages} {031036} (\bibinfo {year}
  {2016})}\BibitemShut {NoStop}%
\bibitem [{\citenamefont {Besse}\ \emph {et~al.}(2020)\citenamefont {Besse}, \citenamefont {Reuer}, \citenamefont {Collodo}, \citenamefont {Wulff}, \citenamefont {Wernli}, \citenamefont {Copetudo}, \citenamefont {Malz}, \citenamefont {Magnard}, \citenamefont {Abdulkadir~Akin}, \citenamefont {Gabureac}, \citenamefont {Norris}, \citenamefont {Cirac}, \citenamefont {Wallraff},\ and\ \citenamefont {Eichler}}]{multi-comm}%
  \BibitemOpen
  \bibfield  {author} {\bibinfo {author} {\bibfnamefont {J.-C.}\ \bibnamefont {Besse}}, \bibinfo {author} {\bibfnamefont {K.}~\bibnamefont {Reuer}}, \bibinfo {author} {\bibfnamefont {M.~C.}\ \bibnamefont {Collodo}}, \bibinfo {author} {\bibfnamefont {A.}~\bibnamefont {Wulff}}, \bibinfo {author} {\bibfnamefont {L.}~\bibnamefont {Wernli}}, \bibinfo {author} {\bibfnamefont {A.}~\bibnamefont {Copetudo}}, \bibinfo {author} {\bibfnamefont {D.}~\bibnamefont {Malz}}, \bibinfo {author} {\bibfnamefont {P.}~\bibnamefont {Magnard}}, \bibinfo {author} {\bibfnamefont {A.}~\bibnamefont {Abdulkadir~Akin}}, \bibinfo {author} {\bibfnamefont {M.}~\bibnamefont {Gabureac}}, \bibinfo {author} {\bibfnamefont {G.~J.}\ \bibnamefont {Norris}}, \bibinfo {author} {\bibfnamefont {J.~I.}\ \bibnamefont {Cirac}}, \bibinfo {author} {\bibfnamefont {A.}~\bibnamefont {Wallraff}},\ and\ \bibinfo {author} {\bibfnamefont {C.}~\bibnamefont {Eichler}},\ }\bibfield  {title} {\bibinfo {title} {Realizing a deterministic source of multipartite-entangled
  photonic qubits},\ }\href@noop {} {\bibfield  {journal} {\bibinfo  {journal} {Nat. Commun.}\ }\textbf {\bibinfo {volume} {11}},\ \bibinfo {pages} {4877} (\bibinfo {year} {2020})}\BibitemShut {NoStop}%
\bibitem [{\citenamefont {Redchenko}\ \emph {et~al.}(2023)\citenamefont {Redchenko}, \citenamefont {Poshakinskiy}, \citenamefont {Sett}, \citenamefont {Žemlička}, \citenamefont {Poddubny},\ and\ \citenamefont {Fink}}]{wqed-scat}%
  \BibitemOpen
  \bibfield  {author} {\bibinfo {author} {\bibfnamefont {E.~S.}\ \bibnamefont {Redchenko}}, \bibinfo {author} {\bibfnamefont {A.~V.}\ \bibnamefont {Poshakinskiy}}, \bibinfo {author} {\bibfnamefont {R.}~\bibnamefont {Sett}}, \bibinfo {author} {\bibfnamefont {M.}~\bibnamefont {Žemlička}}, \bibinfo {author} {\bibfnamefont {A.~N.}\ \bibnamefont {Poddubny}},\ and\ \bibinfo {author} {\bibfnamefont {J.~M.}\ \bibnamefont {Fink}},\ }\bibfield  {title} {\bibinfo {title} {{T}unable directional photon scattering from a pair of superconducting qubits},\ }\href@noop {} {\bibfield  {journal} {\bibinfo  {journal} {Nat. Commun.}\ }\textbf {\bibinfo {volume} {14}},\ \bibinfo {pages} {2998} (\bibinfo {year} {2023})}\BibitemShut {NoStop}%
\bibitem [{\citenamefont {Pichler}\ \emph {et~al.}(2015)\citenamefont {Pichler}, \citenamefont {Ramos}, \citenamefont {Daley},\ and\ \citenamefont {Zoller}}]{pichler2015}%
  \BibitemOpen
  \bibfield  {author} {\bibinfo {author} {\bibfnamefont {H.}~\bibnamefont {Pichler}}, \bibinfo {author} {\bibfnamefont {T.}~\bibnamefont {Ramos}}, \bibinfo {author} {\bibfnamefont {A.~J.}\ \bibnamefont {Daley}},\ and\ \bibinfo {author} {\bibfnamefont {P.}~\bibnamefont {Zoller}},\ }\bibfield  {title} {\bibinfo {title} {Quantum optics of chiral spin networks},\ }\href@noop {} {\bibfield  {journal} {\bibinfo  {journal} {Phys. Rev. A}\ }\textbf {\bibinfo {volume} {91}},\ \bibinfo {pages} {042116} (\bibinfo {year} {2015})}\BibitemShut {NoStop}%
\bibitem [{\citenamefont {Ramos}\ \emph {et~al.}(2016)\citenamefont {Ramos}, \citenamefont {Vermersch}, \citenamefont {Hauke}, \citenamefont {Pichler},\ and\ \citenamefont {Zoller}}]{ramos2016}%
  \BibitemOpen
  \bibfield  {author} {\bibinfo {author} {\bibfnamefont {T.}~\bibnamefont {Ramos}}, \bibinfo {author} {\bibfnamefont {B.}~\bibnamefont {Vermersch}}, \bibinfo {author} {\bibfnamefont {P.}~\bibnamefont {Hauke}}, \bibinfo {author} {\bibfnamefont {H.}~\bibnamefont {Pichler}},\ and\ \bibinfo {author} {\bibfnamefont {P.}~\bibnamefont {Zoller}},\ }\bibfield  {title} {\bibinfo {title} {{Non-{M}arkovian dynamics in chiral quantum networks with spins and photons}},\ }\href@noop {} {\bibfield  {journal} {\bibinfo  {journal} {Phys. Rev. A}\ }\textbf {\bibinfo {volume} {93}},\ \bibinfo {pages} {062104} (\bibinfo {year} {2016})}\BibitemShut {NoStop}%
\bibitem [{\citenamefont {Lodahl}\ \emph {et~al.}(2017)\citenamefont {Lodahl}, \citenamefont {Mahmoodian}, \citenamefont {Stobbe}, \citenamefont {Rauschenbeutel}, \citenamefont {Schneeweiss}, \citenamefont {Volz}, \citenamefont {Pichler},\ and\ \citenamefont {Zoller}}]{lodahl2017chiral}%
  \BibitemOpen
  \bibfield  {author} {\bibinfo {author} {\bibfnamefont {P.}~\bibnamefont {Lodahl}}, \bibinfo {author} {\bibfnamefont {S.}~\bibnamefont {Mahmoodian}}, \bibinfo {author} {\bibfnamefont {S.}~\bibnamefont {Stobbe}}, \bibinfo {author} {\bibfnamefont {A.}~\bibnamefont {Rauschenbeutel}}, \bibinfo {author} {\bibfnamefont {P.}~\bibnamefont {Schneeweiss}}, \bibinfo {author} {\bibfnamefont {J.}~\bibnamefont {Volz}}, \bibinfo {author} {\bibfnamefont {H.}~\bibnamefont {Pichler}},\ and\ \bibinfo {author} {\bibfnamefont {P.}~\bibnamefont {Zoller}},\ }\bibfield  {title} {\bibinfo {title} {Chiral quantum optics},\ }\href@noop {} {\bibfield  {journal} {\bibinfo  {journal} {Nature}\ }\textbf {\bibinfo {volume} {541}},\ \bibinfo {pages} {473} (\bibinfo {year} {2017})}\BibitemShut {NoStop}%
\bibitem [{\citenamefont {Gonzalez-Tudela}\ \emph {et~al.}(2011)\citenamefont {Gonzalez-Tudela}, \citenamefont {Martin-Cano}, \citenamefont {Moreno}, \citenamefont {Martin-Moreno}, \citenamefont {Tejedor},\ and\ \citenamefont {Garcia-Vidal}}]{refereeA1}%
  \BibitemOpen
  \bibfield  {author} {\bibinfo {author} {\bibfnamefont {A.}~\bibnamefont {Gonzalez-Tudela}}, \bibinfo {author} {\bibfnamefont {D.}~\bibnamefont {Martin-Cano}}, \bibinfo {author} {\bibfnamefont {E.}~\bibnamefont {Moreno}}, \bibinfo {author} {\bibfnamefont {L.}~\bibnamefont {Martin-Moreno}}, \bibinfo {author} {\bibfnamefont {C.}~\bibnamefont {Tejedor}},\ and\ \bibinfo {author} {\bibfnamefont {F.~J.}\ \bibnamefont {Garcia-Vidal}},\ }\bibfield  {title} {\bibinfo {title} {{Entanglement of Two Qubits Mediated by One-Dimensional Plasmonic Waveguides}},\ }\href@noop {} {\bibfield  {journal} {\bibinfo  {journal} {Phys. Rev. Lett.}\ }\textbf {\bibinfo {volume} {106}},\ \bibinfo {pages} {020501} (\bibinfo {year} {2011})}\BibitemShut {NoStop}%
\bibitem [{\citenamefont {Shahmoon}\ and\ \citenamefont {Kurizki}(2013)}]{refereeA2}%
  \BibitemOpen
  \bibfield  {author} {\bibinfo {author} {\bibfnamefont {E.}~\bibnamefont {Shahmoon}}\ and\ \bibinfo {author} {\bibfnamefont {G.}~\bibnamefont {Kurizki}},\ }\bibfield  {title} {\bibinfo {title} {Nonradiative interaction and entanglement between distant atoms},\ }\href@noop {} {\bibfield  {journal} {\bibinfo  {journal} {Phys. Rev. A}\ }\textbf {\bibinfo {volume} {87}},\ \bibinfo {pages} {033831} (\bibinfo {year} {2013})}\BibitemShut {NoStop}%
\bibitem [{\citenamefont {Facchi}\ \emph {et~al.}(2016)\citenamefont {Facchi}, \citenamefont {Kim}, \citenamefont {Pascazio}, \citenamefont {Pepe}, \citenamefont {Pomarico},\ and\ \citenamefont {Tufarelli}}]{waveguide_pra}%
  \BibitemOpen
  \bibfield  {author} {\bibinfo {author} {\bibfnamefont {P.}~\bibnamefont {Facchi}}, \bibinfo {author} {\bibfnamefont {M.~S.}\ \bibnamefont {Kim}}, \bibinfo {author} {\bibfnamefont {S.}~\bibnamefont {Pascazio}}, \bibinfo {author} {\bibfnamefont {F.~V.}\ \bibnamefont {Pepe}}, \bibinfo {author} {\bibfnamefont {D.}~\bibnamefont {Pomarico}},\ and\ \bibinfo {author} {\bibfnamefont {T.}~\bibnamefont {Tufarelli}},\ }\bibfield  {title} {\bibinfo {title} {Bound states and entanglement generation in waveguide quantum electrodynamics},\ }\href@noop {} {\bibfield  {journal} {\bibinfo  {journal} {Phys. Rev. A}\ }\textbf {\bibinfo {volume} {94}},\ \bibinfo {pages} {043839} (\bibinfo {year} {2016})}\BibitemShut {NoStop}%
\bibitem [{\citenamefont {Zhang}\ and\ \citenamefont {Baranger}(2019)}]{baranger}%
  \BibitemOpen
  \bibfield  {author} {\bibinfo {author} {\bibfnamefont {X.~H.~H.}\ \bibnamefont {Zhang}}\ and\ \bibinfo {author} {\bibfnamefont {H.~U.}\ \bibnamefont {Baranger}},\ }\bibfield  {title} {\bibinfo {title} {{Heralded Bell State of Dissipative Qubits Using Classical Light in a Waveguide}},\ }\href@noop {} {\bibfield  {journal} {\bibinfo  {journal} {Phys. Rev. Lett.}\ }\textbf {\bibinfo {volume} {122}},\ \bibinfo {pages} {140502} (\bibinfo {year} {2019})}\BibitemShut {NoStop}%
\bibitem [{\citenamefont {Zheng}\ and\ \citenamefont {Baranger}(2013)}]{baranger2013}%
  \BibitemOpen
  \bibfield  {author} {\bibinfo {author} {\bibfnamefont {H.}~\bibnamefont {Zheng}}\ and\ \bibinfo {author} {\bibfnamefont {H.~U.}\ \bibnamefont {Baranger}},\ }\bibfield  {title} {\bibinfo {title} {{Persistent Quantum Beats and Long-Distance Entanglement from Waveguide-Mediated Interactions}},\ }\href@noop {} {\bibfield  {journal} {\bibinfo  {journal} {Phys. Rev. Lett.}\ }\textbf {\bibinfo {volume} {110}},\ \bibinfo {pages} {113601} (\bibinfo {year} {2013})}\BibitemShut {NoStop}%
\bibitem [{\citenamefont {Gonzalez-Ballestero}\ \emph {et~al.}(2013)\citenamefont {Gonzalez-Ballestero}, \citenamefont {Garcia-Vidal},\ and\ \citenamefont {Moreno}}]{NJP}%
  \BibitemOpen
  \bibfield  {author} {\bibinfo {author} {\bibfnamefont {C.}~\bibnamefont {Gonzalez-Ballestero}}, \bibinfo {author} {\bibfnamefont {F.~J.}\ \bibnamefont {Garcia-Vidal}},\ and\ \bibinfo {author} {\bibfnamefont {E.}~\bibnamefont {Moreno}},\ }\bibfield  {title} {\bibinfo {title} {{Non-Markovian effects in waveguide-mediated entanglement}},\ }\href@noop {} {\bibfield  {journal} {\bibinfo  {journal} {New J. Phys.}\ }\textbf {\bibinfo {volume} {15}},\ \bibinfo {pages} {073015} (\bibinfo {year} {2013})}\BibitemShut {NoStop}%
\bibitem [{\citenamefont {Redchenko}\ and\ \citenamefont {Yudson}(2014)}]{yudson2014}%
  \BibitemOpen
  \bibfield  {author} {\bibinfo {author} {\bibfnamefont {E.~S.}\ \bibnamefont {Redchenko}}\ and\ \bibinfo {author} {\bibfnamefont {V.~I.}\ \bibnamefont {Yudson}},\ }\bibfield  {title} {\bibinfo {title} {Decay of metastable excited states of two qubits in a waveguide},\ }\href@noop {} {\bibfield  {journal} {\bibinfo  {journal} {Phys. Rev. A}\ }\textbf {\bibinfo {volume} {90}},\ \bibinfo {pages} {063829} (\bibinfo {year} {2014})}\BibitemShut {NoStop}%
\bibitem [{\citenamefont {Laakso}\ and\ \citenamefont {Pletyukhov}(2014)}]{laakso}%
  \BibitemOpen
  \bibfield  {author} {\bibinfo {author} {\bibfnamefont {M.}~\bibnamefont {Laakso}}\ and\ \bibinfo {author} {\bibfnamefont {M.}~\bibnamefont {Pletyukhov}},\ }\bibfield  {title} {\bibinfo {title} {{Scattering of Two Photons from Two Distant Qubits: Exact Solution}},\ }\href@noop {} {\bibfield  {journal} {\bibinfo  {journal} {Phys. Rev. Lett.}\ }\textbf {\bibinfo {volume} {113}},\ \bibinfo {pages} {183601} (\bibinfo {year} {2014})}\BibitemShut {NoStop}%
\bibitem [{\citenamefont {van Loo}\ \emph {et~al.}(2013)\citenamefont {van Loo}, \citenamefont {Fedorov}, \citenamefont {Lalumi\`ere}, \citenamefont {Sanders}, \citenamefont {Blais},\ and\ \citenamefont {Wallraff}}]{Fedorov1}%
  \BibitemOpen
  \bibfield  {author} {\bibinfo {author} {\bibfnamefont {A.~F.}\ \bibnamefont {van Loo}}, \bibinfo {author} {\bibfnamefont {A.}~\bibnamefont {Fedorov}}, \bibinfo {author} {\bibfnamefont {K.}~\bibnamefont {Lalumi\`ere}}, \bibinfo {author} {\bibfnamefont {B.~C.}\ \bibnamefont {Sanders}}, \bibinfo {author} {\bibfnamefont {A.}~\bibnamefont {Blais}},\ and\ \bibinfo {author} {\bibfnamefont {A.}~\bibnamefont {Wallraff}},\ }\bibfield  {title} {\bibinfo {title} {{Photon-Mediated Interactions Between Distant Artificial Atoms}},\ }\href@noop {} {\bibfield  {journal} {\bibinfo  {journal} {Science}\ }\textbf {\bibinfo {volume} {342}},\ \bibinfo {pages} {1494} (\bibinfo {year} {2013})}\BibitemShut {NoStop}%
\bibitem [{\citenamefont {Rosario~Hamann}\ \emph {et~al.}(2018)\citenamefont {Rosario~Hamann}, \citenamefont {M\"uller}, \citenamefont {Jerger}, \citenamefont {Zanner}, \citenamefont {Combes}, \citenamefont {Pletyukhov}, \citenamefont {Weides}, \citenamefont {Stace},\ and\ \citenamefont {Fedorov}}]{Fedorov2}%
  \BibitemOpen
  \bibfield  {author} {\bibinfo {author} {\bibfnamefont {A.}~\bibnamefont {Rosario~Hamann}}, \bibinfo {author} {\bibfnamefont {C.}~\bibnamefont {M\"uller}}, \bibinfo {author} {\bibfnamefont {M.}~\bibnamefont {Jerger}}, \bibinfo {author} {\bibfnamefont {M.}~\bibnamefont {Zanner}}, \bibinfo {author} {\bibfnamefont {J.}~\bibnamefont {Combes}}, \bibinfo {author} {\bibfnamefont {M.}~\bibnamefont {Pletyukhov}}, \bibinfo {author} {\bibfnamefont {M.}~\bibnamefont {Weides}}, \bibinfo {author} {\bibfnamefont {T.~M.}\ \bibnamefont {Stace}},\ and\ \bibinfo {author} {\bibfnamefont {A.}~\bibnamefont {Fedorov}},\ }\bibfield  {title} {\bibinfo {title} {{Nonreciprocity Realized with Quantum Nonlinearity}},\ }\href@noop {} {\bibfield  {journal} {\bibinfo  {journal} {Phys. Rev. Lett.}\ }\textbf {\bibinfo {volume} {121}},\ \bibinfo {pages} {123601} (\bibinfo {year} {2018})}\BibitemShut {NoStop}%
\bibitem [{\citenamefont {Mascarenhas}\ \emph {et~al.}(2016)\citenamefont {Mascarenhas}, \citenamefont {Santos}, \citenamefont {Auff\`eves},\ and\ \citenamefont {Gerace}}]{Alexia2016}%
  \BibitemOpen
  \bibfield  {author} {\bibinfo {author} {\bibfnamefont {E.}~\bibnamefont {Mascarenhas}}, \bibinfo {author} {\bibfnamefont {M.~F.}\ \bibnamefont {Santos}}, \bibinfo {author} {\bibfnamefont {A.}~\bibnamefont {Auff\`eves}},\ and\ \bibinfo {author} {\bibfnamefont {D.}~\bibnamefont {Gerace}},\ }\bibfield  {title} {\bibinfo {title} {Quantum rectifier in a one-dimensional photonic channel},\ }\href@noop {} {\bibfield  {journal} {\bibinfo  {journal} {Phys. Rev. A}\ }\textbf {\bibinfo {volume} {93}},\ \bibinfo {pages} {043821} (\bibinfo {year} {2016})}\BibitemShut {NoStop}%
\bibitem [{\citenamefont {Calaj\'o}\ \emph {et~al.}(2019{\natexlab{a}})\citenamefont {Calaj\'o}, \citenamefont {Fang}, \citenamefont {Baranger},\ and\ \citenamefont {Ciccarello}}]{Calajo}%
  \BibitemOpen
  \bibfield  {author} {\bibinfo {author} {\bibfnamefont {G.}~\bibnamefont {Calaj\'o}}, \bibinfo {author} {\bibfnamefont {Y.-L.~L.}\ \bibnamefont {Fang}}, \bibinfo {author} {\bibfnamefont {H.~U.}\ \bibnamefont {Baranger}},\ and\ \bibinfo {author} {\bibfnamefont {F.}~\bibnamefont {Ciccarello}},\ }\bibfield  {title} {\bibinfo {title} {{Exciting a Bound State in the Continuum through Multiphoton Scattering Plus Delayed Quantum Feedback}},\ }\href@noop {} {\bibfield  {journal} {\bibinfo  {journal} {Phys. Rev. Lett.}\ }\textbf {\bibinfo {volume} {122}},\ \bibinfo {pages} {073601} (\bibinfo {year} {2019}{\natexlab{a}})}\BibitemShut {NoStop}%
\bibitem [{\citenamefont {Lonigro}\ \emph {et~al.}(2021{\natexlab{a}})\citenamefont {Lonigro}, \citenamefont {Facchi}, \citenamefont {Greentree}, \citenamefont {Pascazio}, \citenamefont {Pepe},\ and\ \citenamefont {Pomarico}}]{waveguide_pra4}%
  \BibitemOpen
  \bibfield  {author} {\bibinfo {author} {\bibfnamefont {D.}~\bibnamefont {Lonigro}}, \bibinfo {author} {\bibfnamefont {P.}~\bibnamefont {Facchi}}, \bibinfo {author} {\bibfnamefont {A.~D.}\ \bibnamefont {Greentree}}, \bibinfo {author} {\bibfnamefont {S.}~\bibnamefont {Pascazio}}, \bibinfo {author} {\bibfnamefont {F.~V.}\ \bibnamefont {Pepe}},\ and\ \bibinfo {author} {\bibfnamefont {D.}~\bibnamefont {Pomarico}},\ }\bibfield  {title} {\bibinfo {title} {Photon-emitter dressed states in a closed waveguide},\ }\href@noop {} {\bibfield  {journal} {\bibinfo  {journal} {Phys. Rev. A}\ }\textbf {\bibinfo {volume} {104}},\ \bibinfo {pages} {053702} (\bibinfo {year} {2021}{\natexlab{a}})}\BibitemShut {NoStop}%
\bibitem [{\citenamefont {Guimond}\ \emph {et~al.}(2020)\citenamefont {Guimond}, \citenamefont {Vermersch}, \citenamefont {Juan}, \citenamefont {Sharafiev}, \citenamefont {Kirchmair},\ and\ \citenamefont {Zoller}}]{Guimond2020unidirectional}%
  \BibitemOpen
  \bibfield  {author} {\bibinfo {author} {\bibfnamefont {P.-O.}\ \bibnamefont {Guimond}}, \bibinfo {author} {\bibfnamefont {B.}~\bibnamefont {Vermersch}}, \bibinfo {author} {\bibfnamefont {M.}~\bibnamefont {Juan}}, \bibinfo {author} {\bibfnamefont {A.}~\bibnamefont {Sharafiev}}, \bibinfo {author} {\bibfnamefont {G.}~\bibnamefont {Kirchmair}},\ and\ \bibinfo {author} {\bibfnamefont {P.}~\bibnamefont {Zoller}},\ }\bibfield  {title} {\bibinfo {title} {A unidirectional on-chip photonic interface for superconducting circuits},\ }\href@noop {} {\bibfield  {journal} {\bibinfo  {journal} {npj Quantum Information}\ }\textbf {\bibinfo {volume} {6}},\ \bibinfo {pages} {32} (\bibinfo {year} {2020})}\BibitemShut {NoStop}%
\bibitem [{\citenamefont {Scigliuzzo}\ \emph {et~al.}(2022)\citenamefont {Scigliuzzo}, \citenamefont {Calajò}, \citenamefont {Ciccarello}, \citenamefont {Lozano}, \citenamefont {Bengtsson}, \citenamefont {Scarlino}, \citenamefont {Wallraff}, \citenamefont {Chang}, \citenamefont {Delsing},\ and\ \citenamefont {Gasparinetti}}]{gasparinetti}%
  \BibitemOpen
  \bibfield  {author} {\bibinfo {author} {\bibfnamefont {M.}~\bibnamefont {Scigliuzzo}}, \bibinfo {author} {\bibfnamefont {G.}~\bibnamefont {Calajò}}, \bibinfo {author} {\bibfnamefont {F.}~\bibnamefont {Ciccarello}}, \bibinfo {author} {\bibfnamefont {D.~P.}\ \bibnamefont {Lozano}}, \bibinfo {author} {\bibfnamefont {A.}~\bibnamefont {Bengtsson}}, \bibinfo {author} {\bibfnamefont {P.}~\bibnamefont {Scarlino}}, \bibinfo {author} {\bibfnamefont {A.}~\bibnamefont {Wallraff}}, \bibinfo {author} {\bibfnamefont {D.}~\bibnamefont {Chang}}, \bibinfo {author} {\bibfnamefont {P.}~\bibnamefont {Delsing}},\ and\ \bibinfo {author} {\bibfnamefont {S.}~\bibnamefont {Gasparinetti}},\ }\bibfield  {title} {\bibinfo {title} {Controlling atom-photon bound states in an array of josephson-junction resonators},\ }\href@noop {} {\bibfield  {journal} {\bibinfo  {journal} {Phys. Rev. X}\ }\textbf {\bibinfo {volume} {12}},\ \bibinfo {pages} {031036} (\bibinfo {year} {2022})}\BibitemShut {NoStop}%
\bibitem [{\citenamefont {Gasparinetti}(2023)}]{gasparinetti2023photons}%
  \BibitemOpen
  \bibfield  {author} {\bibinfo {author} {\bibfnamefont {S.}~\bibnamefont {Gasparinetti}},\ }\bibfield  {title} {\bibinfo {title} {Photons go one way or another},\ }\href@noop {} {\bibfield  {journal} {\bibinfo  {journal} {Nat. Phys.}\ }\textbf {\bibinfo {volume} {19}},\ \bibinfo {pages} {310} (\bibinfo {year} {2023})}\BibitemShut {NoStop}%
\bibitem [{\citenamefont {Solano}\ \emph {et~al.}(2023)\citenamefont {Solano}, \citenamefont {Barberis-Blostein},\ and\ \citenamefont {Sinha}}]{solano2023}%
  \BibitemOpen
  \bibfield  {author} {\bibinfo {author} {\bibfnamefont {P.}~\bibnamefont {Solano}}, \bibinfo {author} {\bibfnamefont {P.}~\bibnamefont {Barberis-Blostein}},\ and\ \bibinfo {author} {\bibfnamefont {K.}~\bibnamefont {Sinha}},\ }\bibfield  {title} {\bibinfo {title} {Dissimilar collective decay and directional emission from two quantum emitters},\ }\href@noop {} {\bibfield  {journal} {\bibinfo  {journal} {Phys. Rev. A}\ }\textbf {\bibinfo {volume} {107}},\ \bibinfo {pages} {023723} (\bibinfo {year} {2023})}\BibitemShut {NoStop}%
\bibitem [{\citenamefont {Dinc}\ \emph {et~al.}(2019)\citenamefont {Dinc}, \citenamefont {Ercan},\ and\ \citenamefont {Bra\'nczyk}}]{Dinc}%
  \BibitemOpen
  \bibfield  {author} {\bibinfo {author} {\bibfnamefont {F.}~\bibnamefont {Dinc}}, \bibinfo {author} {\bibfnamefont {I.}~\bibnamefont {Ercan}},\ and\ \bibinfo {author} {\bibfnamefont {A.~M.}\ \bibnamefont {Bra\'nczyk}},\ }\bibfield  {title} {\bibinfo {title} {{Exact Markovian and non-Markovian time dynamics in waveguide QED: collective interactions, bound states in continuum, superradiance and subradiance}},\ }\href@noop {} {\bibfield  {journal} {\bibinfo  {journal} {Quantum}\ }\textbf {\bibinfo {volume} {3}},\ \bibinfo {pages} {213} (\bibinfo {year} {2019})}\BibitemShut {NoStop}%
\bibitem [{\citenamefont {Lalumi\`ere}\ \emph {et~al.}(2013)\citenamefont {Lalumi\`ere}, \citenamefont {Sanders}, \citenamefont {van Loo}, \citenamefont {Fedorov}, \citenamefont {Wallraff},\ and\ \citenamefont {Blais}}]{lalumiere2013}%
  \BibitemOpen
  \bibfield  {author} {\bibinfo {author} {\bibfnamefont {K.}~\bibnamefont {Lalumi\`ere}}, \bibinfo {author} {\bibfnamefont {B.~C.}\ \bibnamefont {Sanders}}, \bibinfo {author} {\bibfnamefont {A.~F.}\ \bibnamefont {van Loo}}, \bibinfo {author} {\bibfnamefont {A.}~\bibnamefont {Fedorov}}, \bibinfo {author} {\bibfnamefont {A.}~\bibnamefont {Wallraff}},\ and\ \bibinfo {author} {\bibfnamefont {A.}~\bibnamefont {Blais}},\ }\bibfield  {title} {\bibinfo {title} {{Input-output theory for waveguide QED with an ensemble of inhomogeneous atoms}},\ }\href@noop {} {\bibfield  {journal} {\bibinfo  {journal} {Phys. Rev. A}\ }\textbf {\bibinfo {volume} {88}},\ \bibinfo {pages} {043806} (\bibinfo {year} {2013})}\BibitemShut {NoStop}%
\bibitem [{\citenamefont {Sanchez-Burillo}\ \emph {et~al.}(2017)\citenamefont {Sanchez-Burillo}, \citenamefont {Zueco}, \citenamefont {Martin-Moreno},\ and\ \citenamefont {Garcia-Ripoll}}]{boundstates2017}%
  \BibitemOpen
  \bibfield  {author} {\bibinfo {author} {\bibfnamefont {E.}~\bibnamefont {Sanchez-Burillo}}, \bibinfo {author} {\bibfnamefont {D.}~\bibnamefont {Zueco}}, \bibinfo {author} {\bibfnamefont {L.}~\bibnamefont {Martin-Moreno}},\ and\ \bibinfo {author} {\bibfnamefont {J.~J.}\ \bibnamefont {Garcia-Ripoll}},\ }\bibfield  {title} {\bibinfo {title} {{Dynamical signatures of bound states in waveguide QED}},\ }\href@noop {} {\bibfield  {journal} {\bibinfo  {journal} {Phys. Rev. A}\ }\textbf {\bibinfo {volume} {96}},\ \bibinfo {pages} {023831} (\bibinfo {year} {2017})}\BibitemShut {NoStop}%
\bibitem [{\citenamefont {Tsoi}\ and\ \citenamefont {Law}(2008)}]{leo4}%
  \BibitemOpen
  \bibfield  {author} {\bibinfo {author} {\bibfnamefont {T.~S.}\ \bibnamefont {Tsoi}}\ and\ \bibinfo {author} {\bibfnamefont {C.~K.}\ \bibnamefont {Law}},\ }\bibfield  {title} {\bibinfo {title} {Quantum interference effects of a single photon interacting with an atomic chain inside a one-dimensional waveguide},\ }\href@noop {} {\bibfield  {journal} {\bibinfo  {journal} {Phys. Rev. A}\ }\textbf {\bibinfo {volume} {78}},\ \bibinfo {pages} {063832} (\bibinfo {year} {2008})}\BibitemShut {NoStop}%
\bibitem [{\citenamefont {Bello}\ \emph {et~al.}(2019)\citenamefont {Bello}, \citenamefont {Platero}, \citenamefont {Cirac},\ and\ \citenamefont {Gonz\'alez-Tudela}}]{bello}%
  \BibitemOpen
  \bibfield  {author} {\bibinfo {author} {\bibfnamefont {M.}~\bibnamefont {Bello}}, \bibinfo {author} {\bibfnamefont {G.}~\bibnamefont {Platero}}, \bibinfo {author} {\bibfnamefont {J.~I.}\ \bibnamefont {Cirac}},\ and\ \bibinfo {author} {\bibfnamefont {A.}~\bibnamefont {Gonz\'alez-Tudela}},\ }\bibfield  {title} {\bibinfo {title} {{Unconventional quantum optics in topological waveguide QED}},\ }\href@noop {} {\bibfield  {journal} {\bibinfo  {journal} {Sci. Adv.}\ }\textbf {\bibinfo {volume} {5}},\ \bibinfo {pages} {7} (\bibinfo {year} {2019})}\BibitemShut {NoStop}%
\bibitem [{\citenamefont {Facchi}\ \emph {et~al.}(2019)\citenamefont {Facchi}, \citenamefont {Lonigro}, \citenamefont {Pascazio}, \citenamefont {Pepe},\ and\ \citenamefont {Pomarico}}]{waveguide_pra3}%
  \BibitemOpen
  \bibfield  {author} {\bibinfo {author} {\bibfnamefont {P.}~\bibnamefont {Facchi}}, \bibinfo {author} {\bibfnamefont {D.}~\bibnamefont {Lonigro}}, \bibinfo {author} {\bibfnamefont {S.}~\bibnamefont {Pascazio}}, \bibinfo {author} {\bibfnamefont {F.~V.}\ \bibnamefont {Pepe}},\ and\ \bibinfo {author} {\bibfnamefont {D.}~\bibnamefont {Pomarico}},\ }\bibfield  {title} {\bibinfo {title} {Bound states in the continuum for an array of quantum emitters},\ }\href@noop {} {\bibfield  {journal} {\bibinfo  {journal} {Phys. Rev. A}\ }\textbf {\bibinfo {volume} {100}},\ \bibinfo {pages} {023834} (\bibinfo {year} {2019})}\BibitemShut {NoStop}%
\bibitem [{\citenamefont {Dong}\ \emph {et~al.}(2021)\citenamefont {Dong}, \citenamefont {Lee},\ and\ \citenamefont {Choi}}]{dong}%
  \BibitemOpen
  \bibfield  {author} {\bibinfo {author} {\bibfnamefont {Y.}~\bibnamefont {Dong}}, \bibinfo {author} {\bibfnamefont {Y.-S.}\ \bibnamefont {Lee}},\ and\ \bibinfo {author} {\bibfnamefont {K.~S.}\ \bibnamefont {Choi}},\ }\bibfield  {title} {\bibinfo {title} {{Waveguide QED toolboxes for synthetic quantum matter with neutral atoms}},\ }\href@noop {} {\bibfield  {journal} {\bibinfo  {journal} {Phys. Rev. A}\ }\textbf {\bibinfo {volume} {104}},\ \bibinfo {pages} {053703} (\bibinfo {year} {2021})}\BibitemShut {NoStop}%
\bibitem [{\citenamefont {Fang}\ \emph {et~al.}(2014)\citenamefont {Fang}, \citenamefont {Zheng},\ and\ \citenamefont {Baranger}}]{fang14}%
  \BibitemOpen
  \bibfield  {author} {\bibinfo {author} {\bibfnamefont {Y.~L.~L.}\ \bibnamefont {Fang}}, \bibinfo {author} {\bibfnamefont {H.}~\bibnamefont {Zheng}},\ and\ \bibinfo {author} {\bibfnamefont {H.~U.}\ \bibnamefont {Baranger}},\ }\bibfield  {title} {\bibinfo {title} {One-dimensional waveguide coupled to multiple qubits: photon-photon correlations},\ }\href@noop {} {\bibfield  {journal} {\bibinfo  {journal} {EPJ Quantum Technol.}\ }\textbf {\bibinfo {volume} {1}},\ \bibinfo {pages} {3} (\bibinfo {year} {2014})}\BibitemShut {NoStop}%
\bibitem [{\citenamefont {Gu}\ \emph {et~al.}(2017)\citenamefont {Gu}, \citenamefont {Kockum}, \citenamefont {Miranowicz}, \citenamefont {Liu},\ and\ \citenamefont {Nori}}]{gu}%
  \BibitemOpen
  \bibfield  {author} {\bibinfo {author} {\bibfnamefont {X.}~\bibnamefont {Gu}}, \bibinfo {author} {\bibfnamefont {A.~F.}\ \bibnamefont {Kockum}}, \bibinfo {author} {\bibfnamefont {A.}~\bibnamefont {Miranowicz}}, \bibinfo {author} {\bibfnamefont {Y.-X.}\ \bibnamefont {Liu}},\ and\ \bibinfo {author} {\bibfnamefont {F.}~\bibnamefont {Nori}},\ }\bibfield  {title} {\bibinfo {title} {Microwave photonics with superconducting quantum circuits},\ }\href@noop {} {\bibfield  {journal} {\bibinfo  {journal} {Phys. Rep.}\ }\textbf {\bibinfo {volume} {718}},\ \bibinfo {pages} {1} (\bibinfo {year} {2017})}\BibitemShut {NoStop}%
\bibitem [{\citenamefont {Guimond}\ \emph {et~al.}(2016)\citenamefont {Guimond}, \citenamefont {Pichler}, \citenamefont {Rauschenbeutel},\ and\ \citenamefont {Zoller}}]{guimond}%
  \BibitemOpen
  \bibfield  {author} {\bibinfo {author} {\bibfnamefont {P.}~\bibnamefont {Guimond}}, \bibinfo {author} {\bibfnamefont {H.}~\bibnamefont {Pichler}}, \bibinfo {author} {\bibfnamefont {A.}~\bibnamefont {Rauschenbeutel}},\ and\ \bibinfo {author} {\bibfnamefont {P.}~\bibnamefont {Zoller}},\ }\bibfield  {title} {\bibinfo {title} {Chiral quantum optics with v-level atoms and coherent quantum feedback},\ }\href@noop {} {\bibfield  {journal} {\bibinfo  {journal} {Phys. Rev. A}\ }\textbf {\bibinfo {volume} {94}},\ \bibinfo {pages} {033829} (\bibinfo {year} {2016})}\BibitemShut {NoStop}%
\bibitem [{\citenamefont {Paulisch}\ \emph {et~al.}(2016)\citenamefont {Paulisch}, \citenamefont {Kimble},\ and\ \citenamefont {Gonz\'alez-Tudela}}]{paulisch}%
  \BibitemOpen
  \bibfield  {author} {\bibinfo {author} {\bibfnamefont {V.}~\bibnamefont {Paulisch}}, \bibinfo {author} {\bibfnamefont {H.}~\bibnamefont {Kimble}},\ and\ \bibinfo {author} {\bibfnamefont {A.}~\bibnamefont {Gonz\'alez-Tudela}},\ }\bibfield  {title} {\bibinfo {title} {Universal quantum computation in waveguide {QED} using decoherence free subspaces},\ }\href@noop {} {\bibfield  {journal} {\bibinfo  {journal} {New J. Phys.}\ }\textbf {\bibinfo {volume} {18}},\ \bibinfo {pages} {043041} (\bibinfo {year} {2016})}\BibitemShut {NoStop}%
\bibitem [{\citenamefont {Calaj\'o}\ \emph {et~al.}(2016)\citenamefont {Calaj\'o}, \citenamefont {Ciccarello}, \citenamefont {Chang},\ and\ \citenamefont {Rabl}}]{calajo15}%
  \BibitemOpen
  \bibfield  {author} {\bibinfo {author} {\bibfnamefont {G.}~\bibnamefont {Calaj\'o}}, \bibinfo {author} {\bibfnamefont {F.}~\bibnamefont {Ciccarello}}, \bibinfo {author} {\bibfnamefont {D.}~\bibnamefont {Chang}},\ and\ \bibinfo {author} {\bibfnamefont {P.}~\bibnamefont {Rabl}},\ }\bibfield  {title} {\bibinfo {title} {{Atom-field dressed states in slow-light waveguide QED}},\ }\href@noop {} {\bibfield  {journal} {\bibinfo  {journal} {Phys. Rev. A}\ }\textbf {\bibinfo {volume} {93}},\ \bibinfo {pages} {033833} (\bibinfo {year} {2016})}\BibitemShut {NoStop}%
\bibitem [{\citenamefont {Kockum}\ \emph {et~al.}(2018)\citenamefont {Kockum}, \citenamefont {Johansson},\ and\ \citenamefont {Nori}}]{Kockum}%
  \BibitemOpen
  \bibfield  {author} {\bibinfo {author} {\bibfnamefont {A.~F.}\ \bibnamefont {Kockum}}, \bibinfo {author} {\bibfnamefont {G.}~\bibnamefont {Johansson}},\ and\ \bibinfo {author} {\bibfnamefont {F.}~\bibnamefont {Nori}},\ }\bibfield  {title} {\bibinfo {title} {{Decoherence-Free Interaction between Giant Atoms in Waveguide Quantum Electrodynamics}},\ }\href@noop {} {\bibfield  {journal} {\bibinfo  {journal} {Phys. Rev. Lett.}\ }\textbf {\bibinfo {volume} {120}},\ \bibinfo {pages} {140404} (\bibinfo {year} {2018})}\BibitemShut {NoStop}%
\bibitem [{\citenamefont {Goban}\ \emph {et~al.}(2015)\citenamefont {Goban}, \citenamefont {Hung}, \citenamefont {Hood}, \citenamefont {Yu}, \citenamefont {Muniz}, \citenamefont {Painter},\ and\ \citenamefont {Kimble}}]{kimble2015}%
  \BibitemOpen
  \bibfield  {author} {\bibinfo {author} {\bibfnamefont {A.}~\bibnamefont {Goban}}, \bibinfo {author} {\bibfnamefont {C.~L.}\ \bibnamefont {Hung}}, \bibinfo {author} {\bibfnamefont {J.~D.}\ \bibnamefont {Hood}}, \bibinfo {author} {\bibfnamefont {S.~P.}\ \bibnamefont {Yu}}, \bibinfo {author} {\bibfnamefont {J.~A.}\ \bibnamefont {Muniz}}, \bibinfo {author} {\bibfnamefont {O.}~\bibnamefont {Painter}},\ and\ \bibinfo {author} {\bibfnamefont {H.~J.}\ \bibnamefont {Kimble}},\ }\bibfield  {title} {\bibinfo {title} {{Superradiance for Atoms Trapped along a Photonic Crystal Waveguide}},\ }\href@noop {} {\bibfield  {journal} {\bibinfo  {journal} {Phys. Rev. Lett.}\ }\textbf {\bibinfo {volume} {115}},\ \bibinfo {pages} {063601} (\bibinfo {year} {2015})}\BibitemShut {NoStop}%
\bibitem [{\citenamefont {Lonigro}\ \emph {et~al.}(2021{\natexlab{b}})\citenamefont {Lonigro}, \citenamefont {Facchi}, \citenamefont {Pascazio}, \citenamefont {Pepe},\ and\ \citenamefont {Pomarico}}]{waveguide_njp}%
  \BibitemOpen
  \bibfield  {author} {\bibinfo {author} {\bibfnamefont {D.}~\bibnamefont {Lonigro}}, \bibinfo {author} {\bibfnamefont {P.}~\bibnamefont {Facchi}}, \bibinfo {author} {\bibfnamefont {S.}~\bibnamefont {Pascazio}}, \bibinfo {author} {\bibfnamefont {F.~V.}\ \bibnamefont {Pepe}},\ and\ \bibinfo {author} {\bibfnamefont {D.}~\bibnamefont {Pomarico}},\ }\bibfield  {title} {\bibinfo {title} {Stationary excitation waves and multimerization in arrays of quantum emitters},\ }\href@noop {} {\bibfield  {journal} {\bibinfo  {journal} {New J. Phys.}\ }\textbf {\bibinfo {volume} {23}},\ \bibinfo {pages} {103033} (\bibinfo {year} {2021}{\natexlab{b}})}\BibitemShut {NoStop}%
\bibitem [{\citenamefont {Facchi}\ \emph {et~al.}(2018{\natexlab{a}})\citenamefont {Facchi}, \citenamefont {Pascazio}, \citenamefont {Pepe},\ and\ \citenamefont {Yuasa}}]{oscillators}%
  \BibitemOpen
  \bibfield  {author} {\bibinfo {author} {\bibfnamefont {P.}~\bibnamefont {Facchi}}, \bibinfo {author} {\bibfnamefont {S.}~\bibnamefont {Pascazio}}, \bibinfo {author} {\bibfnamefont {F.~V.}\ \bibnamefont {Pepe}},\ and\ \bibinfo {author} {\bibfnamefont {K.}~\bibnamefont {Yuasa}},\ }\bibfield  {title} {\bibinfo {title} {Long-lived entanglement of two multilevel atoms in a waveguide},\ }\href@noop {} {\bibfield  {journal} {\bibinfo  {journal} {J. Phys. Commun.}\ }\textbf {\bibinfo {volume} {2}},\ \bibinfo {pages} {035006} (\bibinfo {year} {2018}{\natexlab{a}})}\BibitemShut {NoStop}%
\bibitem [{\citenamefont {Ramos}\ \emph {et~al.}(2014)\citenamefont {Ramos}, \citenamefont {Pichler}, \citenamefont {Daley},\ and\ \citenamefont {Zoller}}]{ramos14}%
  \BibitemOpen
  \bibfield  {author} {\bibinfo {author} {\bibfnamefont {T.}~\bibnamefont {Ramos}}, \bibinfo {author} {\bibfnamefont {H.}~\bibnamefont {Pichler}}, \bibinfo {author} {\bibfnamefont {A.}~\bibnamefont {Daley}},\ and\ \bibinfo {author} {\bibfnamefont {P.}~\bibnamefont {Zoller}},\ }\bibfield  {title} {\bibinfo {title} {{Quantum Spin Dimers from Chiral Dissipation in Cold-Atom Chains}},\ }\href@noop {} {\bibfield  {journal} {\bibinfo  {journal} {Phys. Rev. Lett.}\ }\textbf {\bibinfo {volume} {113}},\ \bibinfo {pages} {237203} (\bibinfo {year} {2014})}\BibitemShut {NoStop}%
\bibitem [{\citenamefont {Bernien}\ \emph {et~al.}(2017)\citenamefont {Bernien}, \citenamefont {Schwartz}, \citenamefont {Keesling}, \citenamefont {Levine}, \citenamefont {Omran}, \citenamefont {Pichler}, \citenamefont {Choi}, \citenamefont {Zibrov}, \citenamefont {Endres}, \citenamefont {Greiner}, \citenamefont {Vuleti\'c},\ and\ \citenamefont {Lukin}}]{bernien}%
  \BibitemOpen
  \bibfield  {author} {\bibinfo {author} {\bibfnamefont {H.}~\bibnamefont {Bernien}}, \bibinfo {author} {\bibfnamefont {S.}~\bibnamefont {Schwartz}}, \bibinfo {author} {\bibfnamefont {A.}~\bibnamefont {Keesling}}, \bibinfo {author} {\bibfnamefont {H.}~\bibnamefont {Levine}}, \bibinfo {author} {\bibfnamefont {A.}~\bibnamefont {Omran}}, \bibinfo {author} {\bibfnamefont {H.}~\bibnamefont {Pichler}}, \bibinfo {author} {\bibfnamefont {S.}~\bibnamefont {Choi}}, \bibinfo {author} {\bibfnamefont {A.~S.}\ \bibnamefont {Zibrov}}, \bibinfo {author} {\bibfnamefont {M.}~\bibnamefont {Endres}}, \bibinfo {author} {\bibfnamefont {M.}~\bibnamefont {Greiner}}, \bibinfo {author} {\bibfnamefont {V.}~\bibnamefont {Vuleti\'c}},\ and\ \bibinfo {author} {\bibfnamefont {M.~D.}\ \bibnamefont {Lukin}},\ }\bibfield  {title} {\bibinfo {title} {Probing many-body dynamics on a 51-atom quantum simulator},\ }\href@noop {} {\bibfield  {journal} {\bibinfo  {journal} {Nature}\ }\textbf {\bibinfo {volume} {551}},\ \bibinfo {pages} {579}
  (\bibinfo {year} {2017})}\BibitemShut {NoStop}%
\bibitem [{\citenamefont {Facchi}\ \emph {et~al.}(2018{\natexlab{b}})\citenamefont {Facchi}, \citenamefont {Pascazio}, \citenamefont {Pepe},\ and\ \citenamefont {Pomarico}}]{waveguide_pra2}%
  \BibitemOpen
  \bibfield  {author} {\bibinfo {author} {\bibfnamefont {P.}~\bibnamefont {Facchi}}, \bibinfo {author} {\bibfnamefont {S.}~\bibnamefont {Pascazio}}, \bibinfo {author} {\bibfnamefont {F.~V.}\ \bibnamefont {Pepe}},\ and\ \bibinfo {author} {\bibfnamefont {D.}~\bibnamefont {Pomarico}},\ }\bibfield  {title} {\bibinfo {title} {Correlated photon emission by two excited atoms in a waveguide},\ }\href@noop {} {\bibfield  {journal} {\bibinfo  {journal} {Phys. Rev. A}\ }\textbf {\bibinfo {volume} {98}},\ \bibinfo {pages} {063823} (\bibinfo {year} {2018}{\natexlab{b}})}\BibitemShut {NoStop}%
\bibitem [{\citenamefont {Gheeraert}\ \emph {et~al.}(2020)\citenamefont {Gheeraert}, \citenamefont {Kono},\ and\ \citenamefont {Nakamura}}]{Nakamura2020}%
  \BibitemOpen
  \bibfield  {author} {\bibinfo {author} {\bibfnamefont {N.}~\bibnamefont {Gheeraert}}, \bibinfo {author} {\bibfnamefont {S.}~\bibnamefont {Kono}},\ and\ \bibinfo {author} {\bibfnamefont {Y.}~\bibnamefont {Nakamura}},\ }\bibfield  {title} {\bibinfo {title} {Programmable directional emitter and receiver of itinerant microwave photons in a waveguide},\ }\href@noop {} {\bibfield  {journal} {\bibinfo  {journal} {Phys. Rev. A}\ }\textbf {\bibinfo {volume} {102}},\ \bibinfo {pages} {053720} (\bibinfo {year} {2020})}\BibitemShut {NoStop}%
\bibitem [{\citenamefont {Kannan}\ \emph {et~al.}(2023)\citenamefont {Kannan}, \citenamefont {Almanakly}, \citenamefont {Sung}, \citenamefont {Di~Paolo}, \citenamefont {Rower}, \citenamefont {Braum{\"u}ller}, \citenamefont {Melville}, \citenamefont {Niedzielski}, \citenamefont {Karamlou}, \citenamefont {Serniak} \emph {et~al.}}]{kannan2023demand}%
  \BibitemOpen
  \bibfield  {author} {\bibinfo {author} {\bibfnamefont {B.}~\bibnamefont {Kannan}}, \bibinfo {author} {\bibfnamefont {A.}~\bibnamefont {Almanakly}}, \bibinfo {author} {\bibfnamefont {Y.}~\bibnamefont {Sung}}, \bibinfo {author} {\bibfnamefont {A.}~\bibnamefont {Di~Paolo}}, \bibinfo {author} {\bibfnamefont {D.~A.}\ \bibnamefont {Rower}}, \bibinfo {author} {\bibfnamefont {J.}~\bibnamefont {Braum{\"u}ller}}, \bibinfo {author} {\bibfnamefont {A.}~\bibnamefont {Melville}}, \bibinfo {author} {\bibfnamefont {B.~M.}\ \bibnamefont {Niedzielski}}, \bibinfo {author} {\bibfnamefont {A.}~\bibnamefont {Karamlou}}, \bibinfo {author} {\bibfnamefont {K.}~\bibnamefont {Serniak}}, \emph {et~al.},\ }\bibfield  {title} {\bibinfo {title} {On-demand directional microwave photon emission using waveguide quantum electrodynamics},\ }\href@noop {} {\bibfield  {journal} {\bibinfo  {journal} {Nature Physics}\ }\textbf {\bibinfo {volume} {19}},\ \bibinfo {pages} {394} (\bibinfo {year} {2023})}\BibitemShut {NoStop}%
\bibitem [{\citenamefont {Kannan}\ \emph {et~al.}(2020)\citenamefont {Kannan}, \citenamefont {Campbell}, \citenamefont {Vasconcelos}, \citenamefont {Winik}, \citenamefont {Kim}, \citenamefont {Kjaergaard}, \citenamefont {Krantz}, \citenamefont {Melville}, \citenamefont {Niedzielski}, \citenamefont {Yoder} \emph {et~al.}}]{kannan2020generating}%
  \BibitemOpen
  \bibfield  {author} {\bibinfo {author} {\bibfnamefont {B.}~\bibnamefont {Kannan}}, \bibinfo {author} {\bibfnamefont {D.~L.}\ \bibnamefont {Campbell}}, \bibinfo {author} {\bibfnamefont {F.}~\bibnamefont {Vasconcelos}}, \bibinfo {author} {\bibfnamefont {R.}~\bibnamefont {Winik}}, \bibinfo {author} {\bibfnamefont {D.}~\bibnamefont {Kim}}, \bibinfo {author} {\bibfnamefont {M.}~\bibnamefont {Kjaergaard}}, \bibinfo {author} {\bibfnamefont {P.}~\bibnamefont {Krantz}}, \bibinfo {author} {\bibfnamefont {A.}~\bibnamefont {Melville}}, \bibinfo {author} {\bibfnamefont {B.~M.}\ \bibnamefont {Niedzielski}}, \bibinfo {author} {\bibfnamefont {J.}~\bibnamefont {Yoder}}, \emph {et~al.},\ }\bibfield  {title} {\bibinfo {title} {Generating spatially entangled itinerant photons with waveguide quantum electrodynamics},\ }\href@noop {} {\bibfield  {journal} {\bibinfo  {journal} {Science advances}\ }\textbf {\bibinfo {volume} {6}},\ \bibinfo {pages} {eabb8780} (\bibinfo {year} {2020})}\BibitemShut {NoStop}%
\bibitem [{\citenamefont {Merkel}\ and\ \citenamefont {Wilhelm}(2010)}]{merkel2010generation}%
  \BibitemOpen
  \bibfield  {author} {\bibinfo {author} {\bibfnamefont {S.~T.}\ \bibnamefont {Merkel}}\ and\ \bibinfo {author} {\bibfnamefont {F.~K.}\ \bibnamefont {Wilhelm}},\ }\bibfield  {title} {\bibinfo {title} {{Generation and detection of NOON states in superconducting circuits}},\ }\href@noop {} {\bibfield  {journal} {\bibinfo  {journal} {New J. Phys.}\ }\textbf {\bibinfo {volume} {12}},\ \bibinfo {pages} {093036} (\bibinfo {year} {2010})}\BibitemShut {NoStop}%
\bibitem [{\citenamefont {Wang}\ \emph {et~al.}(2011)\citenamefont {Wang}, \citenamefont {Mariantoni}, \citenamefont {Bialczak}, \citenamefont {Lenander}, \citenamefont {Lucero}, \citenamefont {Neeley}, \citenamefont {O'Connell}, \citenamefont {Sank}, \citenamefont {Weides}, \citenamefont {Wenner}, \citenamefont {Yamamoto}, \citenamefont {Yin}, \citenamefont {Zhao}, \citenamefont {Martinis},\ and\ \citenamefont {Cleland}}]{wang2011deterministic}%
  \BibitemOpen
  \bibfield  {author} {\bibinfo {author} {\bibfnamefont {H.}~\bibnamefont {Wang}}, \bibinfo {author} {\bibfnamefont {M.}~\bibnamefont {Mariantoni}}, \bibinfo {author} {\bibfnamefont {R.~C.}\ \bibnamefont {Bialczak}}, \bibinfo {author} {\bibfnamefont {M.}~\bibnamefont {Lenander}}, \bibinfo {author} {\bibfnamefont {E.}~\bibnamefont {Lucero}}, \bibinfo {author} {\bibfnamefont {M.}~\bibnamefont {Neeley}}, \bibinfo {author} {\bibfnamefont {A.~D.}\ \bibnamefont {O'Connell}}, \bibinfo {author} {\bibfnamefont {D.}~\bibnamefont {Sank}}, \bibinfo {author} {\bibfnamefont {M.}~\bibnamefont {Weides}}, \bibinfo {author} {\bibfnamefont {J.}~\bibnamefont {Wenner}}, \bibinfo {author} {\bibfnamefont {T.}~\bibnamefont {Yamamoto}}, \bibinfo {author} {\bibfnamefont {Y.}~\bibnamefont {Yin}}, \bibinfo {author} {\bibfnamefont {J.}~\bibnamefont {Zhao}}, \bibinfo {author} {\bibfnamefont {J.~M.}\ \bibnamefont {Martinis}},\ and\ \bibinfo {author} {\bibfnamefont {A.~N.}\ \bibnamefont {Cleland}},\ }\bibfield  {title} {\bibinfo {title}
  {{Deterministic Entanglement of Photons in Two Superconducting Microwave Resonators}},\ }\href@noop {} {\bibfield  {journal} {\bibinfo  {journal} {Phys. Rev. Lett.}\ }\textbf {\bibinfo {volume} {106}},\ \bibinfo {pages} {060401} (\bibinfo {year} {2011})}\BibitemShut {NoStop}%
\bibitem [{\citenamefont {Lang}\ \emph {et~al.}(2013)\citenamefont {Lang}, \citenamefont {Eichler}, \citenamefont {Steffen}, \citenamefont {Fink}, \citenamefont {Woolley}, \citenamefont {Blais},\ and\ \citenamefont {Wallraff}}]{lang2013correlations}%
  \BibitemOpen
  \bibfield  {author} {\bibinfo {author} {\bibfnamefont {C.}~\bibnamefont {Lang}}, \bibinfo {author} {\bibfnamefont {C.}~\bibnamefont {Eichler}}, \bibinfo {author} {\bibfnamefont {L.}~\bibnamefont {Steffen}}, \bibinfo {author} {\bibfnamefont {J.~M.}\ \bibnamefont {Fink}}, \bibinfo {author} {\bibfnamefont {M.~J.}\ \bibnamefont {Woolley}}, \bibinfo {author} {\bibfnamefont {A.}~\bibnamefont {Blais}},\ and\ \bibinfo {author} {\bibfnamefont {A.}~\bibnamefont {Wallraff}},\ }\bibfield  {title} {\bibinfo {title} {{Correlations, indistinguishability and entanglement in Hong–Ou–Mandel experiments at microwave frequencies}},\ }\href@noop {} {\bibfield  {journal} {\bibinfo  {journal} {{Nat. Phys.}}\ }\textbf {\bibinfo {volume} {9}},\ \bibinfo {pages} {345} (\bibinfo {year} {2013})}\BibitemShut {NoStop}%
\bibitem [{\citenamefont {Hua}\ \emph {et~al.}(2020)\citenamefont {Hua}, \citenamefont {Tao},\ and\ \citenamefont {Deng}}]{hua2020efficient}%
  \BibitemOpen
  \bibfield  {author} {\bibinfo {author} {\bibfnamefont {M.}~\bibnamefont {Hua}}, \bibinfo {author} {\bibfnamefont {M.-J.}\ \bibnamefont {Tao}},\ and\ \bibinfo {author} {\bibfnamefont {F.-G.}\ \bibnamefont {Deng}},\ }\bibfield  {title} {\bibinfo {title} {{Efficient generation of NOON states on two microwave-photon resonators}},\ }\href@noop {} {\bibfield  {journal} {\bibinfo  {journal} {{Chin. Sci. Bull.}}\ }\textbf {\bibinfo {volume} {59}},\ \bibinfo {pages} {2829} (\bibinfo {year} {2020})}\BibitemShut {NoStop}%
\bibitem [{\citenamefont {Mohseni}\ \emph {et~al.}(2020)\citenamefont {Mohseni}, \citenamefont {Saeidian}, \citenamefont {Dowling},\ and\ \citenamefont {Navarrete-Benlloch}}]{mohseni2020deterministic}%
  \BibitemOpen
  \bibfield  {author} {\bibinfo {author} {\bibfnamefont {N.}~\bibnamefont {Mohseni}}, \bibinfo {author} {\bibfnamefont {S.}~\bibnamefont {Saeidian}}, \bibinfo {author} {\bibfnamefont {J.~P.}\ \bibnamefont {Dowling}},\ and\ \bibinfo {author} {\bibfnamefont {C.}~\bibnamefont {Navarrete-Benlloch}},\ }\bibfield  {title} {\bibinfo {title} {{Deterministic generation of hybrid high-$N$ NOON states with Rydberg atoms trapped in microwave cavities}},\ }\href@noop {} {\bibfield  {journal} {\bibinfo  {journal} {{Phys. Rev. A}}\ }\textbf {\bibinfo {volume} {101}},\ \bibinfo {pages} {013804} (\bibinfo {year} {2020})}\BibitemShut {NoStop}%
\bibitem [{\citenamefont {Su}\ \emph {et~al.}(2013)\citenamefont {Su}, \citenamefont {Yang},\ and\ \citenamefont {Zheng}}]{su2013fast}%
  \BibitemOpen
  \bibfield  {author} {\bibinfo {author} {\bibfnamefont {Q.-P.}\ \bibnamefont {Su}}, \bibinfo {author} {\bibfnamefont {C.-P.}\ \bibnamefont {Yang}},\ and\ \bibinfo {author} {\bibfnamefont {S.-B.}\ \bibnamefont {Zheng}},\ }\bibfield  {title} {\bibinfo {title} {{Fast and simple scheme for generating NOON states of photons in circuit QED}},\ }\href@noop {} {\bibfield  {journal} {\bibinfo  {journal} {{Sci. Rep.}}\ }\textbf {\bibinfo {volume} {4}},\ \bibinfo {pages} {3898} (\bibinfo {year} {2013})}\BibitemShut {NoStop}%
\bibitem [{\citenamefont {Yudson}\ and\ \citenamefont {Reineker}(2008)}]{yudson2008}%
  \BibitemOpen
  \bibfield  {author} {\bibinfo {author} {\bibfnamefont {V.~I.}\ \bibnamefont {Yudson}}\ and\ \bibinfo {author} {\bibfnamefont {P.}~\bibnamefont {Reineker}},\ }\bibfield  {title} {\bibinfo {title} {Multiphoton scattering in a one-dimensional waveguide with resonant atoms},\ }\href@noop {} {\bibfield  {journal} {\bibinfo  {journal} {Phys. Rev. A}\ }\textbf {\bibinfo {volume} {78}},\ \bibinfo {pages} {052713} (\bibinfo {year} {2008})}\BibitemShut {NoStop}%
\bibitem [{\citenamefont {Fang}\ and\ \citenamefont {Baranger}(2015)}]{leo3}%
  \BibitemOpen
  \bibfield  {author} {\bibinfo {author} {\bibfnamefont {Y.~L.~L.}\ \bibnamefont {Fang}}\ and\ \bibinfo {author} {\bibfnamefont {H.~U.}\ \bibnamefont {Baranger}},\ }\bibfield  {title} {\bibinfo {title} {{Waveguide QED: Power spectra and correlations of two photons scattered off multiple distant qubits and a mirror}},\ }\href@noop {} {\bibfield  {journal} {\bibinfo  {journal} {Phys. Rev. A}\ }\textbf {\bibinfo {volume} {91}},\ \bibinfo {pages} {053845} (\bibinfo {year} {2015})}\BibitemShut {NoStop}%
\bibitem [{\citenamefont {Shi}\ \emph {et~al.}(2015)\citenamefont {Shi}, \citenamefont {Chang},\ and\ \citenamefont {Cirac}}]{cirac2015}%
  \BibitemOpen
  \bibfield  {author} {\bibinfo {author} {\bibfnamefont {T.}~\bibnamefont {Shi}}, \bibinfo {author} {\bibfnamefont {D.~E.}\ \bibnamefont {Chang}},\ and\ \bibinfo {author} {\bibfnamefont {J.~I.}\ \bibnamefont {Cirac}},\ }\bibfield  {title} {\bibinfo {title} {Multiphoton-scattering theory and generalized master equations},\ }\href@noop {} {\bibfield  {journal} {\bibinfo  {journal} {Phys. Rev. A}\ }\textbf {\bibinfo {volume} {92}},\ \bibinfo {pages} {053834} (\bibinfo {year} {2015})}\BibitemShut {NoStop}%
\bibitem [{\citenamefont {Gonz\'alez-Tudela}\ \emph {et~al.}(2017)\citenamefont {Gonz\'alez-Tudela}, \citenamefont {Paulisch}, \citenamefont {Kimble},\ and\ \citenamefont {Cirac}}]{paulisch2017}%
  \BibitemOpen
  \bibfield  {author} {\bibinfo {author} {\bibfnamefont {A.}~\bibnamefont {Gonz\'alez-Tudela}}, \bibinfo {author} {\bibfnamefont {V.}~\bibnamefont {Paulisch}}, \bibinfo {author} {\bibfnamefont {H.~J.}\ \bibnamefont {Kimble}},\ and\ \bibinfo {author} {\bibfnamefont {J.~I.}\ \bibnamefont {Cirac}},\ }\bibfield  {title} {\bibinfo {title} {{Efficient Multiphoton Generation in Waveguide Quantum Electrodynamics}},\ }\href@noop {} {\bibfield  {journal} {\bibinfo  {journal} {Phys. Rev. Lett.}\ }\textbf {\bibinfo {volume} {118}},\ \bibinfo {pages} {213601} (\bibinfo {year} {2017})}\BibitemShut {NoStop}%
\bibitem [{\citenamefont {Maffei}\ \emph {et~al.}(2023{\natexlab{a}})\citenamefont {Maffei}, \citenamefont {Goes}, \citenamefont {Wein}, \citenamefont {Jordan}, \citenamefont {Lanco},\ and\ \citenamefont {Auff{\`e}ves}}]{maffei2023energy}%
  \BibitemOpen
  \bibfield  {author} {\bibinfo {author} {\bibfnamefont {M.}~\bibnamefont {Maffei}}, \bibinfo {author} {\bibfnamefont {B.~O.}\ \bibnamefont {Goes}}, \bibinfo {author} {\bibfnamefont {S.~C.}\ \bibnamefont {Wein}}, \bibinfo {author} {\bibfnamefont {A.~N.}\ \bibnamefont {Jordan}}, \bibinfo {author} {\bibfnamefont {L.}~\bibnamefont {Lanco}},\ and\ \bibinfo {author} {\bibfnamefont {A.}~\bibnamefont {Auff{\`e}ves}},\ }\bibfield  {title} {\bibinfo {title} {Energy-efficient quantum non-demolition measurement with a spin-photon interface},\ }\href@noop {} {\bibfield  {journal} {\bibinfo  {journal} {Quantum}\ }\textbf {\bibinfo {volume} {7}},\ \bibinfo {pages} {1099} (\bibinfo {year} {2023}{\natexlab{a}})}\BibitemShut {NoStop}%
\bibitem [{\citenamefont {Maffei}\ \emph {et~al.}(2022)\citenamefont {Maffei}, \citenamefont {Camati},\ and\ \citenamefont {Auff\`eves}}]{Maffei2022Closed}%
  \BibitemOpen
  \bibfield  {author} {\bibinfo {author} {\bibfnamefont {M.}~\bibnamefont {Maffei}}, \bibinfo {author} {\bibfnamefont {P.~A.}\ \bibnamefont {Camati}},\ and\ \bibinfo {author} {\bibfnamefont {A.}~\bibnamefont {Auff\`eves}},\ }\bibfield  {title} {\bibinfo {title} {{Closed-System Solution of the 1D Atom from Collision Model}},\ }\href@noop {} {\bibfield  {journal} {\bibinfo  {journal} {Entropy}\ }\textbf {\bibinfo {volume} {24}},\ \bibinfo {pages} {151} (\bibinfo {year} {2022})}\BibitemShut {NoStop}%
\bibitem [{\citenamefont {Fischer}\ \emph {et~al.}(2018)\citenamefont {Fischer}, \citenamefont {Trivedi}, \citenamefont {Ramasesh}, \citenamefont {Siddiqi},\ and\ \citenamefont {Vu\v{c}kovi\'{c}}}]{Fischer}%
  \BibitemOpen
  \bibfield  {author} {\bibinfo {author} {\bibfnamefont {K.~A.}\ \bibnamefont {Fischer}}, \bibinfo {author} {\bibfnamefont {R.}~\bibnamefont {Trivedi}}, \bibinfo {author} {\bibfnamefont {V.}~\bibnamefont {Ramasesh}}, \bibinfo {author} {\bibfnamefont {I.}~\bibnamefont {Siddiqi}},\ and\ \bibinfo {author} {\bibfnamefont {J.}~\bibnamefont {Vu\v{c}kovi\'{c}}},\ }\bibfield  {title} {\bibinfo {title} {Scattering into one-dimensional waveguides from a coherently-driven quantum-optical system},\ }\href@noop {} {\bibfield  {journal} {\bibinfo  {journal} {{Quantum}}\ }\textbf {\bibinfo {volume} {2}},\ \bibinfo {pages} {69} (\bibinfo {year} {2018})}\BibitemShut {NoStop}%
\bibitem [{\citenamefont {Gardiner}\ and\ \citenamefont {Zoller}(2000)}]{gardinerzoller}%
  \BibitemOpen
  \bibfield  {author} {\bibinfo {author} {\bibfnamefont {C.~W.}\ \bibnamefont {Gardiner}}\ and\ \bibinfo {author} {\bibfnamefont {P.}~\bibnamefont {Zoller}},\ }\href@noop {} {\emph {\bibinfo {title} {{Quantum Noise: A Handbook of Markovian and Non-Markovian Quantum Stochastic Methods with Applications to Quantum Optics}}}}\ (\bibinfo  {publisher} {Springer-Verlag},\ \bibinfo {address} {Berlin-Heidelberg},\ \bibinfo {year} {2000})\BibitemShut {NoStop}%
\bibitem [{\citenamefont {Gardiner}\ and\ \citenamefont {Collett}(1985)}]{Gardiner1985Input}%
  \BibitemOpen
  \bibfield  {author} {\bibinfo {author} {\bibfnamefont {C.~W.}\ \bibnamefont {Gardiner}}\ and\ \bibinfo {author} {\bibfnamefont {M.~J.}\ \bibnamefont {Collett}},\ }\bibfield  {title} {\bibinfo {title} {Input and output in damped quantum systems: Quantum stochastic differential equations and the master equation},\ }\href@noop {} {\bibfield  {journal} {\bibinfo  {journal} {Phys. Rev. A}\ }\textbf {\bibinfo {volume} {31}},\ \bibinfo {pages} {3761} (\bibinfo {year} {1985})}\BibitemShut {NoStop}%
\bibitem [{\citenamefont {Cohen-Tannoudji}\ \emph {et~al.}(1998)\citenamefont {Cohen-Tannoudji}, \citenamefont {Dupont-Roc},\ and\ \citenamefont {Grynberg}}]{cohentannoudjiAPI}%
  \BibitemOpen
  \bibfield  {author} {\bibinfo {author} {\bibfnamefont {C.}~\bibnamefont {Cohen-Tannoudji}}, \bibinfo {author} {\bibfnamefont {J.}~\bibnamefont {Dupont-Roc}},\ and\ \bibinfo {author} {\bibfnamefont {G.}~\bibnamefont {Grynberg}},\ }\href@noop {} {\emph {\bibinfo {title} {{A}tom-{P}hoton {I}nteractions: {B}asic {P}rocesses and {A}pplications}}}\ (\bibinfo  {publisher} {Wiley-VCH Verlag GmbH},\ \bibinfo {address} {Weinheim},\ \bibinfo {year} {1998})\BibitemShut {NoStop}%
\bibitem [{\citenamefont {Burgarth}\ \emph {et~al.}(2023)\citenamefont {Burgarth}, \citenamefont {Facchi}, \citenamefont {Hillier},\ and\ \citenamefont {Ligab{\`o}}}]{burgarth2023taming}%
  \BibitemOpen
  \bibfield  {author} {\bibinfo {author} {\bibfnamefont {D.}~\bibnamefont {Burgarth}}, \bibinfo {author} {\bibfnamefont {P.}~\bibnamefont {Facchi}}, \bibinfo {author} {\bibfnamefont {R.}~\bibnamefont {Hillier}},\ and\ \bibinfo {author} {\bibfnamefont {M.}~\bibnamefont {Ligab{\`o}}},\ }\bibfield  {title} {\bibinfo {title} {Taming the rotating wave approximation},\ }\href@noop {} {\bibfield  {journal} {\bibinfo  {journal} {arXiv preprint arXiv:2301.02269}\ } (\bibinfo {year} {2023})}\BibitemShut {NoStop}%
\bibitem [{\citenamefont {Combes}\ \emph {et~al.}(2017)\citenamefont {Combes}, \citenamefont {Kerckhoff},\ and\ \citenamefont {Sarovar}}]{combes_slh_2017}%
  \BibitemOpen
  \bibfield  {author} {\bibinfo {author} {\bibfnamefont {J.}~\bibnamefont {Combes}}, \bibinfo {author} {\bibfnamefont {J.}~\bibnamefont {Kerckhoff}},\ and\ \bibinfo {author} {\bibfnamefont {M.}~\bibnamefont {Sarovar}},\ }\bibfield  {title} {\bibinfo {title} {The {SLH} framework for modeling quantum input-output networks},\ }\href@noop {} {\bibfield  {journal} {\bibinfo  {journal} {Adv. Phys. X}\ }\textbf {\bibinfo {volume} {2}},\ \bibinfo {pages} {784} (\bibinfo {year} {2017})}\BibitemShut {NoStop}%
\bibitem [{\citenamefont {Gorini}\ \emph {et~al.}(1976)\citenamefont {Gorini}, \citenamefont {Kossakowski},\ and\ \citenamefont {Sudarshan}}]{GKS}%
  \BibitemOpen
  \bibfield  {author} {\bibinfo {author} {\bibfnamefont {V.}~\bibnamefont {Gorini}}, \bibinfo {author} {\bibfnamefont {A.}~\bibnamefont {Kossakowski}},\ and\ \bibinfo {author} {\bibfnamefont {E.~C.~G.}\ \bibnamefont {Sudarshan}},\ }\bibfield  {title} {\bibinfo {title} {{Completely positive dynamical semigroups of N-level systems}},\ }\href@noop {} {\bibfield  {journal} {\bibinfo  {journal} {J. Math. Phys.}\ }\textbf {\bibinfo {volume} {17}},\ \bibinfo {pages} {821} (\bibinfo {year} {1976})}\BibitemShut {NoStop}%
\bibitem [{\citenamefont {Lindblad}(1976)}]{L}%
  \BibitemOpen
  \bibfield  {author} {\bibinfo {author} {\bibfnamefont {G.}~\bibnamefont {Lindblad}},\ }\bibfield  {title} {\bibinfo {title} {{On the generators of quantum dynamical semigroups}},\ }\href@noop {} {\bibfield  {journal} {\bibinfo  {journal} {Commun. Math. Phys.}\ }\textbf {\bibinfo {volume} {48}},\ \bibinfo {pages} {119} (\bibinfo {year} {1976})}\BibitemShut {NoStop}%
\bibitem [{\citenamefont {Maffei}\ \emph {et~al.}(2023{\natexlab{b}})\citenamefont {Maffei}, \citenamefont {Pomarico}, \citenamefont {Facchi}, \citenamefont {Magnifico}, \citenamefont {Pascazio},\ and\ \citenamefont {Pepe}}]{SM}%
  \BibitemOpen
  \bibfield  {author} {\bibinfo {author} {\bibfnamefont {M.}~\bibnamefont {Maffei}}, \bibinfo {author} {\bibfnamefont {D.}~\bibnamefont {Pomarico}}, \bibinfo {author} {\bibfnamefont {P.}~\bibnamefont {Facchi}}, \bibinfo {author} {\bibfnamefont {G.}~\bibnamefont {Magnifico}}, \bibinfo {author} {\bibfnamefont {S.}~\bibnamefont {Pascazio}},\ and\ \bibinfo {author} {\bibfnamefont {F.~V.}\ \bibnamefont {Pepe}},\ }\href@noop {} {\bibinfo {title} {Supplemental material}} (\bibinfo {year} {2023}{\natexlab{b}})\BibitemShut {NoStop}%
\bibitem [{\citenamefont {Blow}\ \emph {et~al.}(1990)\citenamefont {Blow}, \citenamefont {Loudon}, \citenamefont {Phoenix},\ and\ \citenamefont {Shepherd}}]{Loudon1990}%
  \BibitemOpen
  \bibfield  {author} {\bibinfo {author} {\bibfnamefont {K.~J.}\ \bibnamefont {Blow}}, \bibinfo {author} {\bibfnamefont {R.}~\bibnamefont {Loudon}}, \bibinfo {author} {\bibfnamefont {S.~J.~D.}\ \bibnamefont {Phoenix}},\ and\ \bibinfo {author} {\bibfnamefont {T.~J.}\ \bibnamefont {Shepherd}},\ }\bibfield  {title} {\bibinfo {title} {Continuum fields in quantum optics},\ }\href@noop {} {\bibfield  {journal} {\bibinfo  {journal} {Phys. Rev. A}\ }\textbf {\bibinfo {volume} {42}},\ \bibinfo {pages} {4102} (\bibinfo {year} {1990})}\BibitemShut {NoStop}%
\bibitem [{\citenamefont {Baragiola}\ \emph {et~al.}(2012)\citenamefont {Baragiola}, \citenamefont {Cook}, \citenamefont {Bra\ifmmode~\acute{n}\else \'{n}\fi{}czyk},\ and\ \citenamefont {Combes}}]{Combes2012PRA}%
  \BibitemOpen
  \bibfield  {author} {\bibinfo {author} {\bibfnamefont {B.~Q.}\ \bibnamefont {Baragiola}}, \bibinfo {author} {\bibfnamefont {R.~L.}\ \bibnamefont {Cook}}, \bibinfo {author} {\bibfnamefont {A.~M.}\ \bibnamefont {Bra\ifmmode~\acute{n}\else \'{n}\fi{}czyk}},\ and\ \bibinfo {author} {\bibfnamefont {J.}~\bibnamefont {Combes}},\ }\bibfield  {title} {\bibinfo {title} {$n$-photon wave packets interacting with an arbitrary quantum system},\ }\href@noop {} {\bibfield  {journal} {\bibinfo  {journal} {Phys. Rev. A}\ }\textbf {\bibinfo {volume} {86}},\ \bibinfo {pages} {013811} (\bibinfo {year} {2012})}\BibitemShut {NoStop}%
\bibitem [{\citenamefont {Pichler}\ and\ \citenamefont {Zoller}(2016)}]{ZollerCM}%
  \BibitemOpen
  \bibfield  {author} {\bibinfo {author} {\bibfnamefont {H.}~\bibnamefont {Pichler}}\ and\ \bibinfo {author} {\bibfnamefont {P.}~\bibnamefont {Zoller}},\ }\bibfield  {title} {\bibinfo {title} {{Photonic Circuits with Time Delays and Quantum Feedback}},\ }\href@noop {} {\bibfield  {journal} {\bibinfo  {journal} {Phys. Rev. Lett.}\ }\textbf {\bibinfo {volume} {116}},\ \bibinfo {pages} {093601} (\bibinfo {year} {2016})}\BibitemShut {NoStop}%
\bibitem [{\citenamefont {Calaj\'o}\ \emph {et~al.}(2019{\natexlab{b}})\citenamefont {Calaj\'o}, \citenamefont {Fang}, \citenamefont {Baranger},\ and\ \citenamefont {Ciccarello}}]{Calajo_scattering}%
  \BibitemOpen
  \bibfield  {author} {\bibinfo {author} {\bibfnamefont {G.}~\bibnamefont {Calaj\'o}}, \bibinfo {author} {\bibfnamefont {Y.-L.~L.}\ \bibnamefont {Fang}}, \bibinfo {author} {\bibfnamefont {H.~U.}\ \bibnamefont {Baranger}},\ and\ \bibinfo {author} {\bibfnamefont {F.}~\bibnamefont {Ciccarello}},\ }\bibfield  {title} {\bibinfo {title} {{Exciting a Bound State in the Continuum through Multiphoton Scattering Plus Delayed Quantum Feedback}},\ }\href@noop {} {\bibfield  {journal} {\bibinfo  {journal} {Phys. Rev. Lett.}\ }\textbf {\bibinfo {volume} {122}},\ \bibinfo {pages} {073601} (\bibinfo {year} {2019}{\natexlab{b}})}\BibitemShut {NoStop}%
\bibitem [{\citenamefont {Guimond}\ \emph {et~al.}(2017)\citenamefont {Guimond}, \citenamefont {Pletyukhov}, \citenamefont {Pichler},\ and\ \citenamefont {Zoller}}]{guimond2017scattering}%
  \BibitemOpen
  \bibfield  {author} {\bibinfo {author} {\bibfnamefont {P.-O.}\ \bibnamefont {Guimond}}, \bibinfo {author} {\bibfnamefont {M.}~\bibnamefont {Pletyukhov}}, \bibinfo {author} {\bibfnamefont {H.}~\bibnamefont {Pichler}},\ and\ \bibinfo {author} {\bibfnamefont {P.}~\bibnamefont {Zoller}},\ }\bibfield  {title} {\bibinfo {title} {Delayed coherent quantum feedback from a scattering theory and a matrix product state perspective},\ }\href@noop {} {\bibfield  {journal} {\bibinfo  {journal} {Quantum Sci. Technol.}\ }\textbf {\bibinfo {volume} {2}},\ \bibinfo {pages} {044012} (\bibinfo {year} {2017})}\BibitemShut {NoStop}%
\end{thebibliography}
\end{document}


\title{SUPPLEMENTAL MATERIAL \\ Directional emission and photon bunching from a qubit pair in waveguide}
\author{Maria Maffei}
\author{Domenico Pomarico}
\author{Paolo Facchi}
\author{\\ Giuseppe Magnifico}
\author{Saverio Pascazio}
\author{Francesco V. Pepe}
\affiliation{Dipartimento di Fisica, Universit\`{a} di Bari, I-70126 Bari, Italy}
\affiliation{INFN, Sezione di Bari, I-70125 Bari, Italy}
\date{\today}
\maketitle

\section{The model}
We consider two identical qubits embedded in a one dimensional waveguide where the electromagnetic field can propagate in both directions. We assume i) rotating wave approximation, ii) flat light-matter coupling, iii) linear dispersion relation. 
The bare qubit Hamiltonian is 
\begin{equation}
H^{(0)}_j= \omega_0 \sigma^{\dagger}_j\sigma_j,	
\end{equation} 
where $\sigma_j=\ket{g_j}\bra{e_j}$, and the qubits are also coupled by the cancellation coupling Hamiltonian
\begin{equation}
	H_c=  J(\sigma_1^+ \sigma_2 + \sigma_1 \sigma_2^+).
\end{equation}
The waveguide is a continuous reservoir of modes of frequency $\omega$ described by the bare Hamiltonian 
\begin{equation}
	H^{(0)}_{f}= \int \dd\omega~\omega \left[a^{\dagger}_{R}(\omega)a_{R}(\omega)+a^{\dagger}_{L}(\omega)a_{L}(\omega)\right],
\end{equation}
where  $[a_{\ell}(\omega),a^{\dagger}_{m}(\omega')]=\delta_{\ell,m}\delta(\omega-\omega')$, with $\ell,m\in\lbrace R,L\rbrace$. The coupling between the qubits and the photon field is
\begin{equation}
	V=  \sqrt{\frac{\gamma}{4\pi}} \int \dd\omega~\left\{\sigma_{1} \left[e^{i \omega_0 x_1/v_g} a^{\dagger}_{R}(\omega)+ e^{-i \omega_0 x_1/v_g} a^{\dagger}_{L}(\omega)\right] + \sigma_{2} \left[e^{i \omega_0 x_2/v_g} a^{\dagger}_{R}(\omega)+ e^{-i \omega_0 x_1/v_g} a^{\dagger}_{L}(\omega)\right]\right\} + H.c.
\end{equation}
with $x_1$ and $x_2$ being the qubits' positions.  By defining the new annihilation operators, or noise operators
\begin{equation}
	b_\ell(t)= \frac{1}{\sqrt{2\pi}} \int \dd\omega e^{-i (\omega-\omega_0)t}a_\ell(\omega),
\end{equation}
  with $[b_{\ell}(t),b^{\dagger}_{m}(t')]=\delta_{\ell,m}\delta(t-t')$,
in the interaction picture with respect to $H^{(0)}=H^{(0)}_1+H^{(0)}_2+H^{(0)}_f$, the light-matter coupling becomes
\begin{align}\label{eq:V_continuousTime_sm}
V_{I}(t)= \sqrt{\frac{\gamma}{2}}\Biggl\lbrace & \sigma_{1} \left[e^{-i \omega_0 x_1/v_g} b^{\dagger}_{R}\left(t-\frac{x_1}{v_g}\right)+ e^{i \omega_0 x_1/v_g} b^{\dagger}_{L}\left(t+\frac{x_1}{v_g}\right)\right]   \nonumber \\ 
& +\sigma_{2} \left[e^{-i \omega_0 x_2/v_g} b^{\dagger}_{R}\left(t-\frac{x_2}{v_g}\right)+ e^{i \omega_0 x_2/v_g} b^{\dagger}_{L}\left(t+\frac{x_2}{v_g}\right)\right] \Biggr\rbrace + \mathrm{H.c.}
\end{align}
By  fixing $x_1=0$ and $x_2=d$, and $\tau=d/v_g$ one gets Hamiltonian~(1) in the text. Under the assumption   $\tau\ll \gamma^{-1}$, it becomes
\begin{align}\label{eq:V_continuousTime2}
V_{I}(t) &= \sqrt{\frac{\gamma}{2}}\Biggl\lbrace  \sigma_{1} \left[b^{\dagger}_{R}\left(t\right)+  b^{\dagger}_{L}\left(t\right)\right]    +\sigma_{2} \left[e^{-i \omega_0 \tau} b^{\dagger}_{R}\left(t\right)+ e^{i \omega_0 \tau} b^{\dagger}_{L}\left(t\right)\right] \Biggr\rbrace + \mathrm{H.c.}
\nonumber\\
&= \sqrt{\frac{\gamma}{2}}\Biggl\lbrace  \left[\sigma_{1} +e^{-i \omega_0 \tau}\sigma_{2} \right] b^{\dagger}_{R}\left(t\right)    + \left[\sigma_{1} +e^{i \omega_0 \tau}\sigma_{2} \right] b^{\dagger}_{L}\left(t\right) \Biggr\rbrace + \mathrm{H.c.}
\end{align}
and the dynamics of the two qubits can  be described by a GKLS master equation~\cite{combesslh2017,GKS,L} as we will now show. 

The noise operators $b_\ell^\dag(t)$ and $b_\ell(t)$ are the central objects of the input-output formalism.
They are rather singular operators that create and destroy a set of continuous bosonic modes, and are the quantum analogues of classical white noise. 
In the interaction picture one can write the Schr\"odinger equation for the unitary propagator of the full system, which is a quantum stochastic differential equation in terms of the quantum noise increment operators 
\begin{equation}
	\mathrm{d}B_{\ell}(t)\equiv \int_{t}^{t+\mathrm{d}t} \mathrm{d}s~b_{\ell}(s), \qquad \mathrm{d}B^\dag_{\ell}(t)\equiv \int_{t}^{t+\mathrm{d}t} \mathrm{d}s~b^\dag_{\ell}(s), \qquad \mathrm{d}\Lambda_{\ell m}(t)\equiv \int_{t}^{t+\mathrm{d}t} \mathrm{d}s~b_{\ell}^\dag(s)b_{m}(s),
\end{equation}
which are the quantum analogues of a Wiener process and satisfy the It\=o table
\begin{equation}
\mathrm{d}B_{\ell}	\mathrm{d}B^\dag_{\ell}= \mathrm{d}t, \qquad
\mathrm{d}B_{\ell} \mathrm{d}\Lambda_{\ell,m} = \mathrm{d}B_{m}, \qquad
\mathrm{d}\Lambda_{\ell m} \mathrm{d}B^\dag_{m}  = \mathrm{d}B^\dag_{m}, \qquad
\mathrm{d}\Lambda_{\ell m} \mathrm{d}\Lambda_{m n}  = \mathrm{d}\Lambda_{\ell n},
\end{equation}
and all the rests are $0$.
The generic form of the quantum stochastic differential equation for the unitary propagator has the form~\cite{Parthasarathy}
\begin{align}
	\label{eq_HP}
    \mathrm{d}U(t) = \Bigg\{&\Big(-i H +\frac{1}{2}\sum_\ell L_\ell^\dag L_\ell  \Big)\, \mathrm{d}t      +\sum_\ell  L_{\ell} \mathrm{d}B^{\dagger}_{\ell} - \sum_{\ell,m}L^{\dagger}_{\ell} S_{\ell m}\mathrm{d}B_{m} + \sum_{\ell,m} \left(S_{\ell m}-\delta_{\ell m}\openone\right)\mathrm{d}\Lambda_{\ell m} \Bigg\}U(t),
\end{align}
where $L_\ell$ is the qubit system operator that couples to the $\ell$th input mode, $H$ is the effective system
Hamiltonian, and $S_{\ell m}$ are system scattering operators. By tracing out the quantum noise fields, one obtains a GKLS master equation for the system with jump operators $L_\ell$.
In our case from the interaction~\eqref{eq:V_continuousTime2} and the It\=o table, one gets~\cite{lalumiere2013,combesslh2017}
\begin{equation}
	H=H_c +\frac{\gamma}{2} (\sigma_1^\dag \sigma_2 + \sigma_1 \sigma_2^\dag), \quad L_R=  -i \sqrt{\frac{\gamma}{2}} \left(\sigma_2 + e^{i\omega_0\tau}\sigma_1 \right), \quad L_L=  -i \sqrt{\frac{\gamma}{2}} \left(\sigma_1 + e^{i\omega_0\tau}\sigma_2 \right),  \quad S_{\ell m}=\delta_{\ell m} e^{i\omega_0\tau} \openone ,
\end{equation}
whence
\begin{align}
	    \mathrm{d}U(t) = \Bigg\{&\Big(-i H +\frac{1}{2}\sum_\ell L_\ell^\dag L_\ell  \Big)\, \mathrm{d}t      +\sum_\ell  \left[ L_{\ell} \mathrm{d}B^{\dagger}_{\ell} - e^{i\omega_0\tau}L^{\dagger}_{\ell} \mathrm{d}B_{\ell} +  \left(e^{i\omega_0\tau}-1\right)\mathrm{d}\Lambda_{\ell \ell} \right]\Bigg\}U(t),
\end{align}
which is the evolution equation~(2) in the text.

\begin{figure}
    \centering
    \includegraphics[width=0.5\columnwidth]{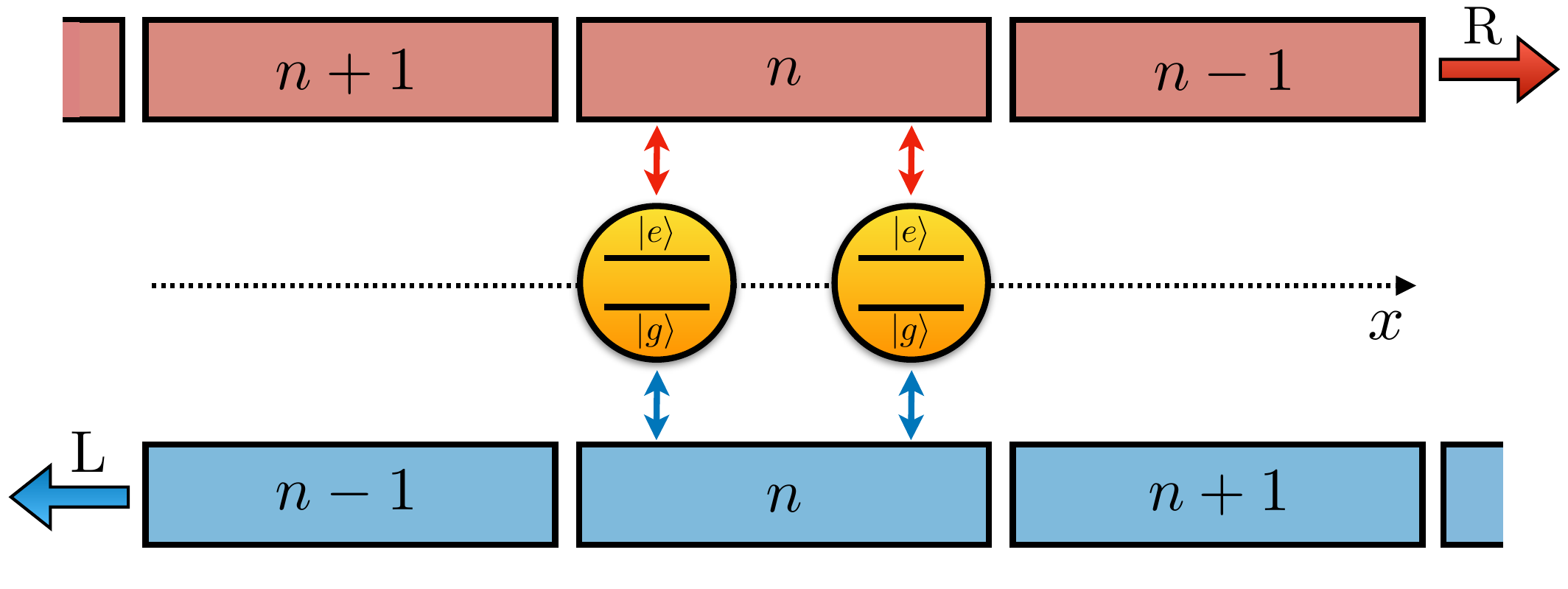}
    \caption{Collision model of the interaction between the qubits and the electromagnetic field propagating in the waveguide. The field is modeled as a stream of bosonic units propagating in opposite directions with velocity $v_g$ and interacting (colliding) with the qubits sequentially. When the time of flight between the qubits $\tau$ is much smaller than the typical time of their spontaneous emission $\gamma^{-1}$, both qubits interact with the same field unit at each collision.}
    \label{fig:panel_1}
\end{figure}

There is an instructive alternative way to derive the unitary propagator, which helps to better understand the physical meaning of approximations
which makes use of the collision model for quantum optics~\cite{Ciccarello2017, Ciccarello2020, ZollerCM}. The time is discretized by using a coarse-graining interval $\Delta t\ll\gamma^{-1}$. Let us notice that, once the inequality $\tau \ll \gamma^{-1}$ is verified, one can set $\tau<\Delta t\ll \gamma^{-1}$. Then, one can define  discrete It\=o increment operators  as 
\begin{equation}
	b_{\ell,n} = \frac{1}{\sqrt{\Delta t}} \int_{n \Delta t}^{(n+1)\Delta t} dt~b_{\ell}(t),
\end{equation}
which satisfy
$[b_{\ell,n},b^{\dagger}_{m,n'}]=\delta_{\ell,m}\delta_{n,n'}$, and the light-matter coupling Eq.~(1) is rewritten as:
\begin{equation}\label{eq:V_discreteTime}
    V_{I}(n)=\frac{1}{\Delta t}\int_{n \Delta t}^{(n+1)\Delta t}dt V_{I}(t) = \frac{i}{\Delta t} \sum_{\ell=R,L} \left( \tilde{\mathcal{J}}^{\,}_{\ell} b^{\dagger}_{\ell,n} - \tilde{\mathcal{J}}^{\dagger}_{\ell} b^{\,}_{\ell,n} \right),
\end{equation}
with 
\begin{align} 
    \tilde{\mathcal{J}}_{R}  = -i \sqrt{\frac{\gamma\Delta t}{2}} \left(\sigma_1 + e^{-i\omega_0\tau}\sigma_2 \right), \qquad  \tilde{\mathcal{J}}_{L} & = -i \sqrt{\frac{\gamma\Delta t}{2}} \left(\sigma_1 + e^{i\omega_0\tau} \sigma_2 \right) . \label{eq:vertexJ2_sm}
\end{align}
being the jump operators associated to right and left emission, respectively. The condition on the coarse graining time $\Delta t>\tau$ implies that, at time $t_n$, both qubits are coupled (collide) with the $n$-th time-bin unit of the propagating field, see Fig.~\ref{fig:panel_1}.

The unitary evolution operator evolving the joint light-matter system during a time interval $[n\Delta t,(n+1)\Delta t]$  is expressed through a Magnus expansion at second order by
\begin{equation}
    U_n = \mathcal{T} \e^{-\ii \int_{n \Delta t}^{(n+1) \Delta t} \dd t \left(V_I(t) + H_c\right)} \approx \mathds{1} - \ii H_c \Delta t - \ii \int_{n \Delta t}^{(n+1) \Delta t} \dd t \ V_I(t) - \frac{1}{2} \int_{n \Delta t}^{(n+1) \Delta t} \dd t \int_{n \Delta t}^{t} \dd t' \ [V_I(t), V_I(t')].
\end{equation}
Let us remark that, by our assumption, the time-of-flight between the emitters $d/\varv_g$ is included in the same time bin. We have that $\int_{n \Delta t}^{(n+1) \Delta t} \dd t \ V_I(t)= V_I(n)\Delta t$, while 
the last term, quadratic in $V_I$, gives an effective exchange interaction between the qubits
\begin{equation}
    H_e = -\frac{\ii}{2\Delta t} \int_{n \Delta t}^{(n+1) \Delta t} \dd t \int_{n \Delta t}^{t} \dd t' \ [V_I(t), V_I(t')] = \frac{\gamma}{2} \sin (\omega_0 \tau) (\sigma_1^\dag \sigma_2 + \sigma_1 \sigma_2^\dag),
\end{equation}
yielding the evolution operator for the $n$th time step:
\begin{equation}\label{eq_Un_sm}
    U_n =\text{exp}\left\lbrace- \frac{i}{\hbar}\Delta t (V_I(n)+ H_{e} + H_c ) \right\rbrace .
\end{equation}
The evolution can then be decomposed in a product, or a sequence of independent collisions, i.e. $\ket{\Psi(t_N)}=\prod_{n=0}^{N-1}U_n\ket{\Psi(0)}$,
which, in the continuous limit $\Delta t \to 0$, yields the stochastic Schr\"odinger equation~(2) of the text.

\section{One-excitation sector}

In order to find the explicit expression of the Kraus operator $\mathcal{K}(t)$ in Eq.~(8), we have to evaluate to continuous time limit, $M\Delta t\rightarrow t$, of the product $\prod_{n=0}^{M-1}\langle 0_{R,n},0_{L,n} \vert U_n \vert 0_{L,n},0_{R,n}\rangle$ with $U_n$ being the expansion of the unitary evolution operator in Eq.~\eqref{eq_Un_sm}, up to the order $\gamma\Delta t$:
\begin{equation}
    U_n \approx \mathds{1} -\frac{\ii}{\hbar} \Delta t (V_I(n) + H'_e) - \frac{\Delta t^2}{2\hbar^2} V_I^2(n).
\end{equation}
We find
\begin{align} \label{eq:s1}
        \bra{0_{R,n}, 0_{L,n} }U_n \ket{0_{L,n}, 0_{R,n}} =& \ 1 - \ii \frac{\gamma}{2} \Delta t\left[ \sin (\omega_0 \tau) -g_c \right](\sigma_1^+ \sigma_2 + \sigma_1 \sigma_2^+) - \frac{\gamma}{2} \Delta t [\sigma_1^+ \sigma_1 + \sigma_2^+ \sigma_2 + \cos (\omega_0 \tau) (\sigma_1^+ \sigma_2 + \sigma_1 \sigma_2^+)] \nonumber \\ 
        = & \ 1 - \frac{\gamma}{2} \Delta t \sum_{s=\pm} \sigma_s^+ \sigma_s (1 + s \e^{\ii \omega_0 \tau} -i s g_c)=1-\frac{\Delta t}{2}\sum_{s=\pm}\mu_s \sigma^{+}_s\sigma_s,
   \end{align}
where we introduced the ladder operators $\sigma_{\pm} = \frac{1}{\sqrt{2}}(\sigma_1 \pm \sigma_2)$ and $\mu_{\pm}=\gamma_{\pm}\pm i\delta$ as defined in the main text. Notice that, in order to derive the above expression, we used
\begin{equation}
    \bra{0_{R,n}, 0_{L,n} } V_I^2(n) \ket{0_{L,n}, 0_{R,n}} = \hbar^2 \frac{\gamma}{\Delta t} [\sigma_1^+ \sigma_1 + \sigma_2^+ \sigma_2 + \cos (\omega_0 \tau) (\sigma_1^+ \sigma_2 + \sigma_1 \sigma_2^{\dagger})].
\end{equation}
At this order of approximation, Eq.~\eqref{eq:s1} is the expansion of an exponential operator:
\begin{align}\label{eq:s2}
        \mathcal{K}(M \Delta t)& \equiv\ \prod_{n=0}^{M-1} \bra{0_{R,n}, 0_{L,n}} U_n \ket{0_{L,n}, 0_{R,n}}= \exp \left(- \frac{M \Delta t}{2}\sum_{s=\pm}\mu_s\sigma^{\dagger}_{s}\sigma_{s} \right) .
\end{align}
By the replacement $M \Delta t\rightarrow t$, one obtains Eq.~(8). 

In order to derive the single-photon amplitudes in Eq.~(10), we must consider that the photon emission can take place during any time bin $[(m-1) \Delta t, m \Delta t]$ with $m\in[0,M]$. The corresponding probability amplitudes read
\begin{equation}\label{eq_amplitude_general}
    \mathcal{A}^{(\psi)}_{R}(t_m)= \bra{1_{R,m}, 0_L , g, g} \prod_{n= 0}^{M-1} U_n \ket{0_R, 0_L, \psi}, \quad \mathcal{A}^{(\psi)}_{L}(t_M)=\bra{0_R, 1_{L,m} , g, g} \prod_{n = 0}^{M-1} U_n \ket{0_R, 0_L, \psi},
\end{equation}
with $\ket{1_{R/L,m}}=b^{\dagger}_{R/L,m}\ket{0_{R/L}}$. In order to evaluate the amplitudes above we decompose them as
\begin{align}
\mathcal{A}^{(\psi)}_{R}(t_m) &=\bra{g,g}\left[\prod_{\ell = m+1}^{M-1}\bra{0_{R,\ell}, 0_{L,\ell} } U_\ell \ket{0_{R,\ell}, 0_{L,\ell}}\right]\bra{1_{R,m}, 0_L } U_{m} \ket{0_{R}, 0_{L}}\left[\prod_{\ell = 0}^{m-1} \bra{0_{R,\ell}, 0_{L,\ell} } U_\ell \ket{0_{R,\ell}, 0_{L,\ell}}\right]\ket{\psi}\\ \nonumber
&= \bra{g,g} \mathcal{K}(t_{M}-m\Delta t) \mathcal{J}_R \mathcal{K}(m\Delta t) \ket{\psi},  
\end{align}
where the emission vertex of a right-moving photon is defined in the main text Eqs.~(3), and similarly for $\mathcal{A}^{(\psi)}_{L}(t_m)$. 

The wavefunctions are obtained by summing the emission amplitudes over all the possible emission times $t_m=m\Delta t$ and taking the continuous time limit $t_m\rightarrow t$, $b_{R/L,m}/\sqrt{\Delta t}\rightarrow b_{R/L}(t) $, and $\sum \Delta t\rightarrow \int dt$:
\begin{align} \label{eq:cont_limit}
\sum_{m=0}^{t_{M}}\mathcal{A}^{(\psi)}_{R/L}(t_m) b^{\dagger}_{R/L,m}\ket{0}\rightarrow\int_{0}^{t}dt'f^{(\psi)}_{R/L}(t')b^{\dagger}_{R/L}(t')\ket{0}.
\end{align}

\subsection{Directional emission}

We can now evaluate explicitly the single-photon amplitudes for initial qubit states $\ket{\phi_{\pm}}= (\ket{e, g} \pm \ii \ket{g, e})/\sqrt{2}$
\begin{align}
    f^{(\phi_{\pm})}_R (t) = & \ \sqrt{\frac{\gamma }{2}} \e^{\mp \ii \pi/4} \e^{-\ii \omega_0 \tau/2} \left( \sin{\left(\frac{\omega_0 \tau}{2}\right)} \e^{- \mu_- t/2} \pm \cos{\left(\frac{\omega_0 \tau}{2}\right)} \e^{- \mu_+ t/2} \right),\\
    f^{(\phi_{\pm})}_L (t) = & \ -\sqrt{\frac{\gamma}{2}} \e^{\mp \ii \pi/4} \e^{\ii \omega_0 \tau/2} \left( \sin{\left(\frac{\omega_0 \tau}{2}\right)} \e^{-\mu_- t/2} \mp \cos{\left(\frac{\omega_0 \tau}{2}\right)} \e^{- \mu_+ t/2} \right),
\end{align}
To obtain the probability $\mathcal{P}^{(\phi_\pm)}_{L(R)}$ that the system emits an $L(R)$-propagating photon during its evolution until the final time we take:
\begin{align}
\mathcal{P}^{(\phi_\pm)}_{L(R)}=\int_{0}^{\infty} dt |f_{L(R)}^{(\phi_\pm)}(t)|^2,
\end{align}
that provides the ratio $r_1(\ket{\psi_L}) = \mathcal{P}^{(\psi_L)}_L/\mathcal{P}^{(\psi_R)}_R$ given in Eq.~(12) in the main text. The ratio $r_1(\ket{eg})$ in Eq.~(13) can be obtained from the single-photon amplitudes $f^{(eg)}_{R/L}(t)=\left[f^{(\phi_{+})}_{R/L}(t)+f^{(\phi_{-})}_{R/L}(t)\right]/\sqrt{2}$.

\section{Two-excitation sector}

Here the initial state is $\ket{0_{R},0_{L},e, e}$, and we consider the probability amplitudes of the trajectories leading the qubits to their ground states through the emission of two photons propagating in the same or in opposite directions. For any couple of emission times $t_{m_1}, t_{m_2}$ with $m_2\geq m_1$ we have:
\begin{align}
    &\mathcal{A}_{R,R}(t_{m_1}, t_{m_2})= \bra{1_{R,m_1},1_{R,m_2}, 0_L , g, g} \prod_{\ell = 0}^{M-1} U_\ell \ket{0_R, 0_L, e, e}, \\
    & \mathcal{A}_{L,L}(t_{m_1}, t_{m_2})=\bra{0_R, 1_{L,m_1},1_{L,m_2} , g, g} \prod_{\ell = 0}^{M-1} U_\ell \ket{0_R, 0_L, e, e}, \\
    &\mathcal{A}_{L,R}(t_{m_1}, t_{m_2})= \bra{1_{R,m_2}, 1_{L,m_1} , g, g} \prod_{\ell = 0}^{M-1} U_\ell \ket{0_R, 0_L, e, e}, \\
    & \mathcal{A}_{R,L}(t_{m_1}, t_{m_2})=\bra{1_{R,m_1},1_{L,m_2} , g, g } \prod_{\ell = 0}^{M-1} U_\ell \ket{0_R, 0_L, e, e}.
\end{align}

We find:
\begin{align}
    \mathcal{A}_{R,R}(t_{m_1}, t_{m_2})&= -\gamma \Delta t \ \e^{\ii \omega_0 \tau} \e^{-\gamma t_{m_1}} \left( \sin{\left(\frac{\omega_0 \tau}{2}\right)}^2 \e^{-\frac{(t_{m_2}-t_{m_1}) \mu_-}{2}} + \cos{\left(\frac{\omega_0 \tau}{2}\right)}^2 \e^{\frac{(t_{m_1}-t_{m_2}) \mu_+}{2}} \right) \\
    \mathcal{A}_{L,L}(t_{m_1}, t_{m_2})&= -\gamma \Delta t \ \e^{-\ii \omega_0 \tau} \e^{-\gamma t_{m_1}} \left( \sin{\left(\frac{\omega_0 \tau}{2}\right)}^2 \e^{-\frac{(t_{m_2}-t_{m_1}) \mu_-}{2}} + \cos{\left(\frac{\omega_0 \tau}{2}\right)}^2 \e^{\frac{(t_{m_1}-t_{m_2}) \mu_+}{2}} \right) \\
    \mathcal{A}_{L,R}(t_{m_1}, t_{m_2})&= \gamma \Delta t \ \e^{-\gamma t_{m_1}} \left( \sin{\left(\frac{\omega_0 \tau}{2}\right)}^2 \e^{-\frac{(t_{m_2}-t_{m_1}) \mu_-}{2}} - \cos{\left(\frac{\omega_0 \tau}{2}\right)}^2 \e^{\frac{(t_{m_1}-t_{m_2}) \mu_+}{2}} \right) = \mathcal{A}_{R,L}(t_{m_1}, t_{m_2}) 
\end{align}

Following the same method applied in the single-excitation sector we compute the probabilities $\mathcal{P}_{j,k}$ (with $j,k=L,R$) by summing $|\mathcal{A}_{j,k}(t_{m_1}, t_{m_2})|^2$ over all the possible emission times $t_{m_1}$ and $t_{m_2}$, then turning the sum into integrals using the limit in Eq.~\eqref{eq:cont_limit}: $\sum_{m_1=0}^{M} \sum_{m_2=0}^{m_1} \mathcal{A}_{\ell, \ell'}(t_{m_1}, t_{m_2}) b_{\ell, m_1}^{\dagger} b_{\ell', m_2}^{\dagger}\ket{0} \rightarrow \int_{0}^{t_{M}} dt_1\int_{0}^{t_1}dt_2 \lambda_{\ell,\ell'}(t_1,t_2) b_{\ell}^{\dagger}(t_1)b_{\ell'}^{\dagger}(t_2) \ket{0}$. We find that the two parallel emission $\mathcal{P}_{LL}$ and $\mathcal{P}_{RR}$ always have the same probability.

We evaluate the fidelity 
\begin{equation}
    F_{\mathrm{N00N}}(t) = \left\lvert \left(\bra{gg}\otimes \bra{\Xi}\right) \ket{\psi(t)} \right\rvert^2
\end{equation}
of the evolved state $\ket{\psi(t)}$ with respect to the asymptotic state $\ket{gg}\otimes \ket{\Xi}$, with $\ket{\Xi}$ the N00N state defined in Eq.~(17) assuming that one of the controlled antiresonance conditions defined in Eq.~(11) holds. Since $\lambda_{L,R}(t,t') = \lambda_{R,L}(t,t') = 0$, we must consider only the contributions $\bra{gg}\otimes\bra{2_R 0_L}\ket{\psi(t)}$ and $\bra{gg}\otimes\bra{0_R 2_L}\ket{\psi(t)}$. The evaluation of the former reads
\begin{align}
    \bra{gg}\otimes\bra{2_R 0_L}\ket{\psi(t)} & = \frac{\gamma}{\sqrt{2}} \int_0^{\infty} \mathrm{d}t_1 \int_0^{\infty} \mathrm{d}t_2 \int_0^{t}\mathrm{d}t_1' \int_0^{t_1'}\mathrm{d}t_2' \ \mathrm{e}^{-\frac{\gamma}{2}(t_1 + t_2)} \lambda_{R,R}(t_1', t_2')  \bra{0_R 0_L} b_R(t_1) b_R(t_2) b_R^\dagger(t_1') b_R^\dagger(t_2') \ket{0_R 0_L} \nonumber \\
    & = -\frac{\mathrm{i}}{\sqrt{2}} (-1)^n (1-\mathrm{e}^{-\gamma t})^2, 
\end{align}
while the latter can be deduced by exploiting the property $\lambda_{L,L}(t,t') = - \lambda_{R,R}(t,t')$. Therefore,
\begin{equation}
    F_{\mathrm{N00N}}(t) = (1-\mathrm{e}^{-\gamma t})^4 \stackrel{ t \rightarrow +\infty} \longrightarrow 1.
\end{equation}
